\documentclass{emulateapj-rtx4}
\usepackage{color}
\usepackage{wrapfig}
\usepackage{lscape}
\usepackage{amsmath}
\usepackage[normalem]{ulem}

\newcommand{\sgra}{Sgr~A*}
\newcommand{\gcrt}{GCRT~J1745$-$3009}
\newcommand{\gcmag}{J1745$-$2900}
\newcommand{\be}{\begin{eqnarray}}
\newcommand{\ee}{\end{eqnarray}}

\newcommand{\arcm}{{}^{\prime}}

\newcommand{\Eone}{{\rm RT\,850630}}
\newcommand{\Etwo}{{\rm RT\,910627}}

\begin{document}
\shorttitle{Transient Events in Archival Very Large Array Observations of the Galactic Center}
\title{Transient Events in Archival Very Large Array Observations of the Galactic Center}

\author{Anirudh Chiti\altaffilmark{1,2,*}, Shami Chatterjee\altaffilmark{1,3}, Robert Wharton\altaffilmark{1}, James Cordes\altaffilmark{1}, T. Joseph W. Lazio\altaffilmark{4}, David L. Kaplan\altaffilmark{5}, Geoffrey C. Bower\altaffilmark{6}, Steve Croft\altaffilmark{7,8}}
\altaffiltext{1}{Department of Astronomy, 
                         Cornell University, 
                         Ithaca, NY 14853, USA}
\altaffiltext{2}{Current Address: Kavli Institute for Astrophysics and 
                         Space Research and Department of Physics,
                         Massachusetts Institute of Technology, 
                         Cambridge, MA 02139, USA}
\altaffiltext{3}{Cornell Center for Astrophysics and Planetary Science,
		         Cornell University,
		         Ithaca, NY 14853, USA}
\altaffiltext{4}{Jet Propulsion Laboratory, 
                      California Institute of Technology, 
                      M/S 138-308, 4800 Oak Grove Dr., 
                      Pasadena, CA 91109, USA}
\altaffiltext{5}{Department of Physics, 
                       University of Wisconsin-Milwaukee, 
                       3135 N Maryland Ave.,  
                       Milwaukee, WI 53201, USA}
\altaffiltext{6}{Academia Sinica Institute for Astronomy 
		      and Astrophysics (ASIAA), 645 N. Aohoku Pl, 
		      Hilo, HI 96720 USA }
\altaffiltext{7}{University of California, Berkeley, Dept of 
                       Astronomy, 501 Campbell Hall \#3411, 
                       Berkeley, CA 94720, USA}
\altaffiltext{8}{Eureka Scientific Inc., 
                      2452 Delmer St Suite 100, 
                      Oakland, CA 94602, USA}
\altaffiltext{*}{Email: \texttt{achiti@mit.edu}}

\begin{abstract}

The Galactic center has some of the highest stellar densities in the Galaxy and a range of 
interstellar scattering properties that may aid in the detection of new 
radio-selected transient events. 
Here we describe a search for radio transients in the Galactic center
using over 200~hours of archival data from the Very Large Array (VLA)
at 5~and~8.4~GHz. Every observation of \sgra\ from 1985$-$2005
has been searched using an automated processing and detection pipeline 
sensitive to transients with timescales between 30 seconds and five
minutes with a typical detection threshold of $\sim$100 mJy. 
Eight possible candidates pass tests 
to filter false-positives from radio-frequency interference, calibration
errors, and imaging artifacts. Two events are identified as promising
candidates based on the smoothness of their light curves. 
Despite the high quality of their light curves, these detections remain suspect due to
evidence of incomplete subtraction of the complex structure in the Galactic center,
and apparent contingency of one detection on reduction routines.
Events of this intensity ($\sim$100\,mJy) and duration ($\sim$100\,s) 
are not obviously associated with known astrophysical sources, and 
no counterparts are found in data at other wavelengths.  
We consider potential sources, including 
Galactic center pulsars, dwarf stars, sources like \gcrt, 
and bursts from X-ray binaries.  None can fully explain
the observed transients, suggesting either a new astrophysical source
or a subtle imaging artifact. More sensitive multiwavelength studies are necessary to 
characterize these events which, if real, occur with a rate of 
$14^{+32}_{-12}~{\rm hr}^{-1}\,{\rm deg}^{-2}$ in the Galactic center.
\end{abstract}


\maketitle

\section{Introduction}
\label{sec:introduction}

A wide variety of astrophysical objects manifest as transient radio sources.
While some objects appear to be transient as a result of propagation effects, 
like the intraday variability of extragalactic sources due to scintillation 
\citep[e.g.,][]{kjw01}, many sources exhibit intrinsic transient radio emission.
Since the emission is changing on short timescales ($\tau \lesssim {\rm days}$), 
radio transients are often associated with compact objects and coherent emission 
processes.  
As a result, studies of short duration radio transients provide a window into
energetic and often unexpected radio emission properties from neutron stars, black
holes, dwarf stars, and planets \citep[e.g.,][]{clm04}.

The public archive of the Very Large Array (VLA) contains high quality data from 
over 30 years of observations, making it an excellent resource for radio transient 
searches.
Many searches for long duration transients on timescales of days to years have 
been conducted in the image domain \citep{bsb+07, bs+11, bfs+11, thw+11, mfo+13}.
These searches look for changes in source flux between different observations, 
so they are most sensitive to transients with durations comparable to the 
time between observations ($\tau \gtrsim {\rm days}$) and have a maximum  
resolution equal to a single observation length ($\tau \sim {\rm hour}$).
Transient timescales of seconds to minutes, however, remain relatively unexplored at radio wavelengths.

Variability on timescales of seconds to minutes is a particularly interesting 
regime in the Galactic center.
As an example, hyperstrong scattering along the line of sight to the Galactic center
may broaden intrinsically narrow pulses to 
$\tau_{\rm sc, \nu} \sim 4\, {\rm s}\,\left(\nu / 5~\rm GHz \right)^{-4}$\citep{lc98}, which is 
comparable to the $10\,\rm s$ sample time of most archival VLA data. Thus, the detection
of transients on these timescales towards the Galactic center can potentially 
constrain the population of giant pulse-emitting pulsars in that region.
Additionally, radio emission from dwarf stars has been previously detected with the VLA
at these timescales \citep{bbb+01, hbl+07, wbz+13}. Given the high stellar density in the
Galactic center, flares from these stars may cause observable transient activity in the region.

In general, the Galactic center is an exciting target for radio transient searches
because the supermassive black hole (\sgra) and very high associated stellar densities can 
lead to astrophysical interactions unlikely to occur anywhere else in the Galaxy.
\citet{mpb+05} have shown that low-mass X-ray binaries are centrally peaked in the 
inner parsec and overabundant with respect to the steep cusp in stellar density, 
indicative of compact objects moving to the Galactic center through dynamical 
friction.  
Theoretical estimates also suggest that as many as $2\times 10^4$ stellar mass black 
holes could reside in the inner parsec as a result of similar processes \citep{meg00}.
Furthermore, previous detections of radio transients in the 
Galactic center \citep{zrg+92, hlk+05, hwl+09, bry+05}   
and a radio-emitting magnetar near \sgra\ 
\citep{mgz+13, sj13, efk+13} suggest that the high density of objects
does enable the detection of novel transient activity.
Regardless of the exact nature of any transient phenomena,
the exploration of radio transients on timescales from seconds to minutes 
presents a useful bridge between previous archival imaging searches 
($\tau \gtrsim {\rm days}$) and more recent fast imaging searches for millisecond 
transients \citep{lbb15}.

\begin{deluxetable*}{cccccccc}  
\tabletypesize{\scriptsize}
\tablewidth{0.99\textwidth}
\tablecolumns{7}
\tablecaption{Summary of VLA data sets analyzed for transient activity}
\tablehead{   
  \colhead{Frequency} &
  \colhead{VLA} &
  \colhead{Projects} &
  \colhead{Time on \sgra} & 
  \colhead{Potential} &
  \colhead{Candidate} &
  \colhead{Resolution} &
  \colhead{Image Dimension}\\
  \colhead{(GHz)} &
  \colhead{Configuration}&
  &
  \colhead{(Hours)}&
  \colhead{Events$^\dagger$} &
  \colhead{Events$^\ddagger$} &
  \colhead{(arcsec)} & 
  \colhead{(arcsec)}
}
\startdata
5  & A  & 57 & 42.70 & 2 & 1 & 0.35 & 72  \\
  & BnA  & 10 & 6.72 & 2 & \nodata & 0.5 & 102 \\
  & B  & 10 & 7.70 & 1 & 1 & 1.0 & 205\\
  & CnB  & 5 & 3.51 & 1 & 1 & 3.0 & 540\\
  & C & 5 & 0.97 & \nodata & \nodata & 3.5 &  540\\
  & DnC & 3 & 9.58 & 2& \nodata & 6.5 & 540 \\
  & D & 9 & 8.44 & 1 & \nodata & 13.0 & 540 \\\\
  
8.4 & A & 45 & 21.64 & 6 & 4 & 0.2 & 82  \\
 & BnA & 22 & 65.11 & 2 & \nodata & 0.5 & 102 \\
 & B & 8 & 3.29 & 1 & \nodata & 0.5 & 102 \\
 & CnB  & 11 & 27.42 & 4 & 1 & 1.5 & 307  \\
 & C & 9 & 2.70 & \nodata & \nodata & 2.0 & 324 \\
 & DnC & 6 & 7.27 & \nodata & \nodata & 4.0 & 324 \\
 & D  & 15 & 7.40 & 1	 & \nodata & 7.0 & 324
\enddata
\tablecomments{A summary of the archival VLA data sets analyzed for transient activity. The columns are: the observation frequency, the array configuration (see section 2.1),  the number of analyzed projects, total integration time on \sgra, plausible transient events that were investigated, and candidate events that passed all tests for validity.\newline
$^\dagger$Potential events failed at least one confirmation test and are referred to as Level 0 candidates in this paper.\newline
$^\ddagger$Candidate events passed the full suite of confirmation tests and are referred to as Level 1 candidates in this paper.}
\label{tab:Archivaldata}
 \end{deluxetable*}

We present the results of an archival VLA search for short-duration 
($\tau \approx 30\,{\rm s} - 5\,{\rm min}$) radio transients 
in the Galactic center. The complete set of archival VLA data  
including \sgra\ from 1985 to 2005 at 5 GHz and 8.4 GHz has been 
searched for short-term variability with a typical detection threshold of 
${\sim}100~\rm mJy$. 
From over 214~hours of on-source time, two promising transient events 
are identified.  
These events pass a rigorous set of tests designed to 
filter out false positives caused by radio frequency interference (RFI), 
but the possibility remains that they are the result of some unknown imaging 
or calibration error.
The rest of the paper is organized as follows.  In Section~\ref{sec:methodology},
we describe the archival data sets used and outline our automated 
processing and transient detection pipeline.  In Section~\ref{sec:results},
the most promising results are identified and discussed.  
The occurrence rates for the observed transients are estimated in 
Section~\ref{sec:Rates} and the possible astrophysical origins are 
discussed in Section~\ref{sec:origin}.

\section{Methodology}
\label{sec:methodology}

\subsection{VLA Data Sets}

\begin{figure}[!ht]
\centering
\includegraphics[width = 0.5\textwidth]{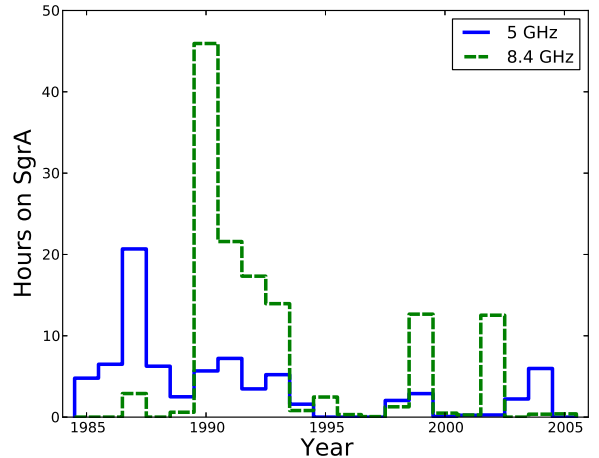}
\caption{Integration time on \sgra\, in 5 GHz and 8.4 GHz observations per year from 1985 to 2005.} 
\label{fig:cadences}
\end{figure}

For this project, we considered all observations of \sgra\ conducted with 
the VLA from 1985 to 2005.  
Observations earlier than 1985 were excluded because past experience suggests the 
early data are unreliable and 
observations later than 2005 were excluded because the array was being upgraded 
to the Expanded Very Large Array.
Table~\ref{tab:Archivaldata} 
summarizes the archival coverage of observations on \sgra. 
The data selected for our search consists of over 214~hours of integration 
time on \sgra\ spanning 215 projects. Each project typically has at least one 
sustained observation of the Galactic center, with only 8 projects 
having less than 5~minutes of integration time on \sgra.
Most of the observations at both 5~GHz and 8.4~GHz were conducted before 
1993 (Figure~\ref{fig:cadences}), meaning 
that any detected radio transients likely occurred over 20 years ago.

\begin{figure*}[ht]
\centering
\centerline{\includegraphics[width = 0.75\textwidth]{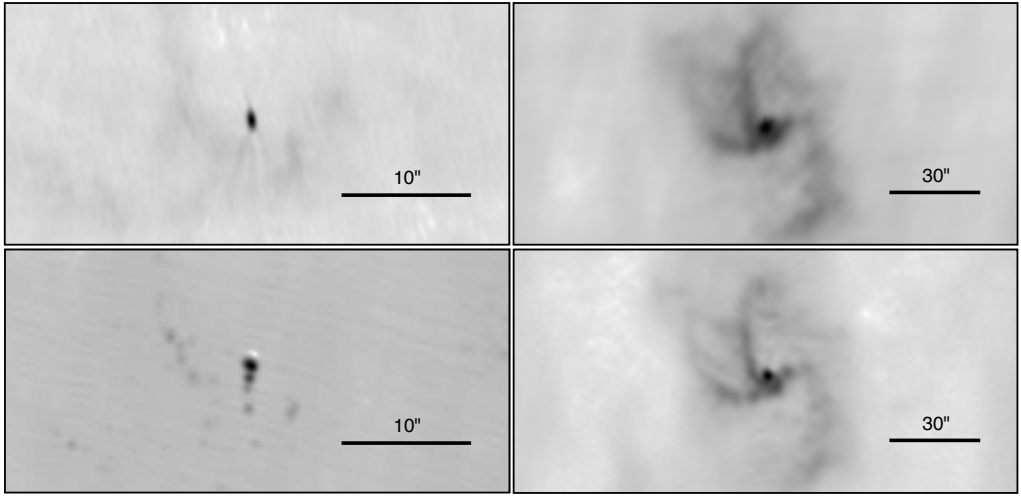}}
\caption{Pipeline-generated images of \sgra. Displayed images have
the median image rms value among all images with the same frequency and configuration. 
Images are scaled by a square-root 
transfer function. 
Upper left: 5~GHz dataset in A configuration.
Upper right: 5~GHz dataset in CnB~configuration.
Lower left: 8.4~GHz dataset in A configuration.
Bottom right: 8.4~GHz dataset in CnB~configuration.
The imaged regions are centered on the same coordinates in this figure and North is up and East is left for
each image.} 
\label{fig:samples}
\end{figure*}

\begin{figure*}[ht]
\centering
\includegraphics[width = 0.375\textwidth]{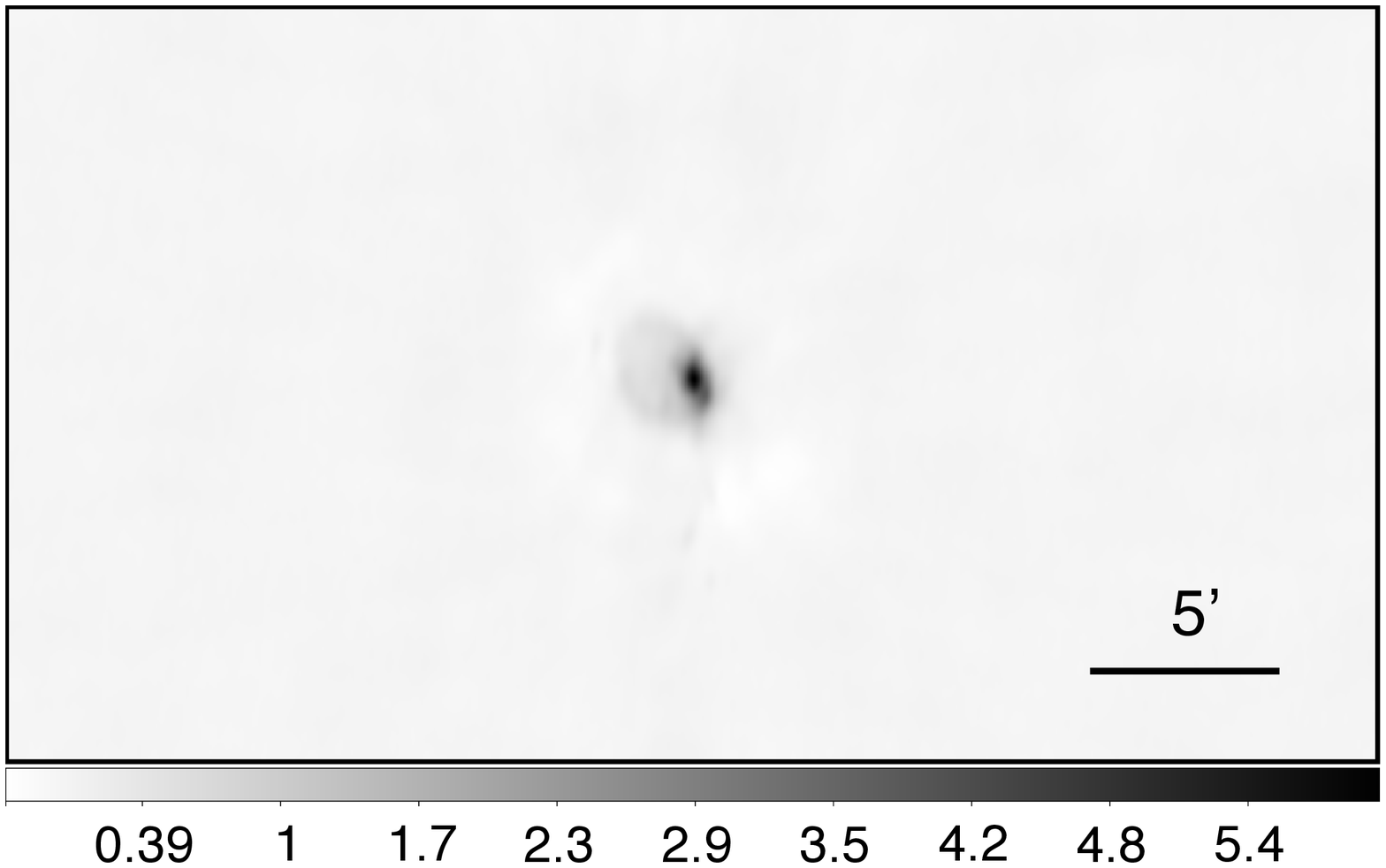}
\includegraphics[width = 0.375\textwidth]{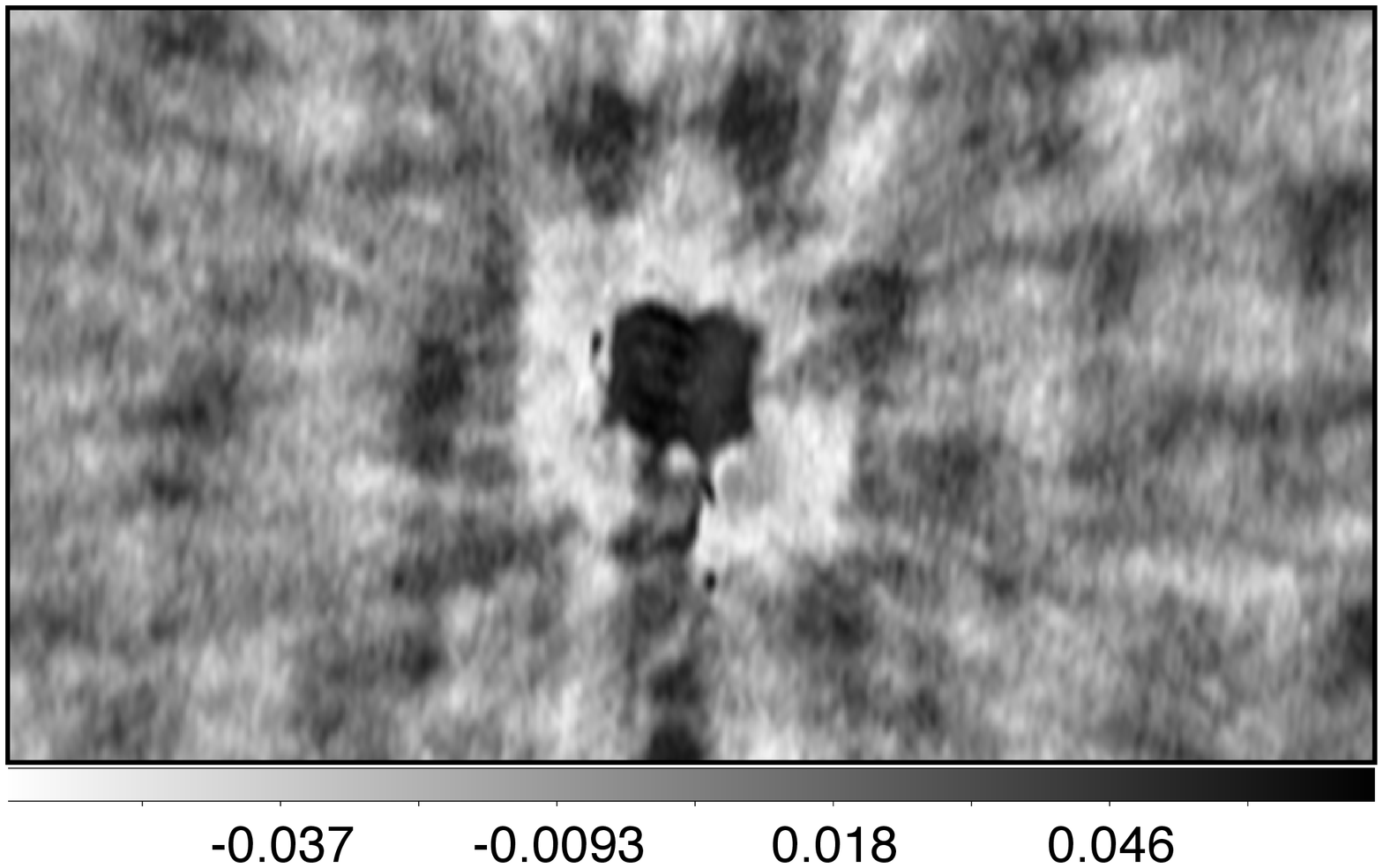}
\caption{A pipeline-calibrated image of \sgra\, from a 5 GHz, D configuration data set (left) and its corresponding model-subtracted image (right). Note that while some structure remains in the model-subtracted image of \sgra, the amplitude of the residual structure is roughly 1\% that of the unsubtracted image. The grey scale is in units of ${\rm Jy\,beam^{-1}}$.}
\label{fig:SampleResidual}
\end{figure*}


The VLA operates in configurations labeled from A to D, where 
A~configuration has the highest resolution and D~configuration
is most sensitive to extended structure.  The VLA also adopts hybrid 
configurations (DnC, CnB, BnA) that deploy an extended
North arm to optimize observations of sources at lower declination. 
Our data spans all configurations of the VLA. At both 5~GHz and 8.4~GHz, 
the majority of observation time was in A, BnA,
or CnB configuration, providing relatively high angular resolution. 
Higher resolution datasets require more processing 
for the same field of view due to the larger number of pixels. 
Instead of imaging the full primary beam (half-power widths of $9\arcm$ and $5.4\arcm$
at 5 and 8 GHz, respectively), we adopted a resolution-dependent
field of view that kept fixed the number of pixels, 
resulting in a minimum area coverage of $72\arcsec\times72\arcsec$ 
in A~configuration.  This requirement guarantees that the most interesting
regions of the Galactic center including the nuclear star cluster, the disk
of young massive stars, and the H{\sc ii} region Sgr~A~West are all included
in every image.
Since all but one potential radio transient investigated in datasets with larger fields of 
view were located within this minimum area, the restriction seems justified in retrospect.

\subsection{Finding Radio Transients}
Using the Common Astronomy Software Applications \citep[CASA;][]{mws+07} 
package\footnote{Available online at 
                 \texttt{http://casa.nrao.edu/}},
we have developed an automated data reduction pipeline to process each 
observation and search for variability on timescales from 30 seconds to 
about five minutes.  While most transient surveys look for variability
across independent observations, we look for variability within observations 
using a CLEAN model subtraction method outlined in \citet{cgb+05}. 

\begin{deluxetable*}{ccccc}[ht!] 
\tablecolumns{5}
\tablewidth{0.99\textwidth}
\tablecaption{RMS statistics of pipeline-generated 10\,s residual images}
\tablehead{   
  \colhead{Frequency} &
  \colhead{VLA Configuration} &
  \colhead{Median RMS} &
  \colhead{Mode RMS} & 
  \colhead{90\% Interval of RMS Values} \\
  \colhead{(GHz)} &
  &
  \colhead{$({\rm mJy\,beam^{-1}})$} &
  \colhead{$({\rm mJy\,beam^{-1}})$} &
  \colhead{$({\rm mJy\,beam^{-1}})$} 
}
\startdata
5   & A & 8.1 & 7.0 & 2.4 -- 15.6\\  
& BnA & 19.4  & 10.7 & 3.7 -- 34.1\\
& B & 18.6  & 17.2  & 8.9 -- 36.2 \\
& CnB & 21.8  & 21.7  & 13.4 -- 43.1 \\
& C & 31.1  & 31.0  & 21.4 -- 42.7 \\
& DnC  & 52.7  & 46.4  & 35.6 -- 93.6 \\
& D  & 49.8  & 50.8  & 17.5 -- 112.7 \\\\

8.4 & A & 6.4 & 5.0 & 2.3 -- 15.6   \\ 
& BnA & 11.1 & 6.8 & 4.8 -- 32.3 \\
& B & 10.5 & 7.8 & 5.7 -- 17.0 \\
& CnB & 18.8  & 12.6 & 10.4 -- 48.5 \\
& C & 18.3 & 10.8 & 10.8 -- 56.4 \\
& DnC & 31.7 & 19.1 & 18.5 -- 52.5 \\
& D & 71.8 & 27.7 & 10.1 -- 211.9
\enddata
\tablecomments{The mean, median, and 90\% interval of RMS values of 10\,s residual images are shown. Cases of extremely low RMS values were excluded from the statistics as potential amplitude mis-calibrations}
\label{tab:rms}
\end{deluxetable*}

\subsubsection{Flagging, Calibration, and Imaging}
\label{sssec:imaging}

For each observation, the data are flagged to remove
corruption from RFI, calibrated using the available flux and phase calibrators, 
then imaged using CLEAN deconvolution.  Though the specific parameters
may change for different observing frequencies and array configurations,
the basic data reduction procedure is as follows.

First, the data set for an observation is retrieved from the NRAO Data 
Archive.\footnote{Available online at \texttt{http://archive.nrao.edu}}  
Before the data can be calibrated,
we run an automated RFI flagging routine using the {TFCrop} algorithm
in the CASA task \texttt{flagdata}.  {TFCrop} is able to identify RFI in an 
uncalibrated data set by fitting a piecewise third-order polynomial to the 
time-averaged bandpass of subintervals of data collected on each baseline.  
The data set is flattened by dividing by the average bandshape for each of the 
considered subintervals.  RFI is identified as outlier data
in either time or frequency.  In addition to RFI flagging, the first 20~seconds
of each scan is flagged to remove any data taken as the telescope was 
settling after slewing back from a calibrator source.

After flagging, the data are calibrated using both the standard gain calibration
procedures and several rounds of self-calibration on \sgra.  For complex
gain calibration, solutions are determined using the calibrator sources provided
in each observation.  After gain calibration, the data are self-calibrated
using \sgra\ itself as a calibrator.  To do this, we image the visibility data using
the Cotton-Schwab implementation of the CLEAN deconvolution algorithm 
\citep{schwab84} with natural weighting of the visibility data.  The deconvolution 
produces a model of the data that is used as the calibrator model in the next
iteration of self-calibration.  Two iterations of phase-only self-calibration
are performed,
followed by one iteration of phase and amplitude self-calibration.  The result of
this process is full-observation model of \sgra\ and a deconvolved image.

Once a model image has been generated for a full observation, the model
is Fourier transformed and subtracted from the calibrated visibility
data.  The resulting model-subtracted visibility data is  
imaged on shorter intervals to search for transient events.
Since one of the fundamental assumptions in interferometric imaging is that 
the sky intensity distribution is constant over the observing span, the generated 
CLEAN model will essentially be the time-averaged intensity 
distribution.  Radio transients that last for a small fraction of the observing
time will only lose a small fraction of their flux to the model.  However, if a radio
transient lasts for a large fraction of the observing time, it will have most of its 
flux contained in the model and be difficult to detect once the model is removed.
As a result, our transient search will be limited to events with durations much shorter
than the full observing span or events that are sufficiently faint to avoid being 
included in the CLEAN model.

A few typical images of the Galactic center produced by the processing pipeline are 
illustrated in Figure~\ref{fig:samples}, and an example images 
of the Galactic center before and after model subtraction are shown in 
Figure~\ref{fig:SampleResidual}.

\begin{deluxetable*}{ccccccccc} 
\tablewidth{0.99\textwidth}
\tabletypesize{\scriptsize}
\tablecolumns{9}
\tablecaption{A summary of the candidate events}
\tablehead{   
  \colhead{Name} &
  \colhead{Freq} &
  \colhead{VLA} &
  \colhead{Duration} & 
  \colhead{$\Delta$RA} &
  \colhead{$\Delta$Dec} &
  \colhead{$\delta \theta_{\rm {\sgra}}$}&
  \colhead{Int. Flux$^\dagger$}&
  \colhead{Peak Flux}\\
  &
  \colhead{(GHz)} &
  \colhead{Config}&
  \colhead{(seconds)}&
  \colhead{(sec)}&
  \colhead{(arcsec)}&
  \colhead{(arcsec)}&
  \colhead{(mJy)}&
  \colhead{$({\rm mJy\,beam^{-1}})$}
}
\startdata
\multicolumn{9}{c}{Level 2}\\
\hline
\noalign{\vskip 2mm} 
RT\,850630 & 5   & CnB  &  120 & 40.00(2)   & $-$23.6(2) & 4.69(21)   & 456 $\pm$ 42  & 149 $\pm$ 26 \\
RT\,910627 & 8.4 & A    &  100  & 38.8578(9) & $-$19.34(2)  & 18.09(2) & 180 $\pm$ 21  & 100 $\pm$ 11\\
\noalign{\vskip 2mm} 
\hline
\noalign{\vskip 1mm} 
\multicolumn{9}{c}{Level 1}\\
\hline
\noalign{\vskip 2mm} 
RT\,910817 & 8.4 & A    &  60  & 40.059(2)  & $-$26.96(4)    & 1.24(4)  & 115 $\pm$ 17  & 72 $\pm$ 7 \\
RT\,950721 & 8.4 & A    &  60  & 40.0469(9) & $-$28.35(3)  & 0.23(2)  & 36  $\pm$ 3   & 55 $\pm$ 5 \\
RT\,921210 & 8.4 & A    &  480 &  40.048(1) & $-$28.30(4)    & 0.18(3)  & 33  $\pm$ 5   & 44 $\pm$ 6 \\
RT\,920208 & 8.4 & CnB  &  140 & 40.13(1)     & $-$26.9(2)   & 1.56(15)   & 100 $\pm$ 12  & 86 $\pm$ 10\\
RT\,860417 & 5   & A    &  50  & 40.582(4)  & $-$22.36(9)  & 8.99(7)  & 180 $\pm$ 30  & 47 $\pm$ 8 \\
RT\,871129 & 5   & B    &  150 & 39.592(4)  & $-$85.1(1)   & 57.23(13)  & 308 $\pm$ 24  & 92 $\pm$ 7 \\
\noalign{\vskip 2mm} 
\hline
\noalign{\vskip 1mm} 
\multicolumn{9}{c}{Level 0}\\
\hline
\noalign{\vskip 2mm} 
RT\,910912 & 8.4 & A    &  150 & 41.8284(6) & $-$32.87(2)  & 23.66(1) & 162 $\pm$ 8   & 192 $\pm$ 10\\
RT\,930109 & 8.4 & A    &  180 & 39.6448(4) & $-$17.45(1)  & 12.05 (1)& 116 $\pm$ 4   & 99  $\pm$ 4 \\
RT\,900704 & 8.4 & BnA  &  340 & 40.092(7)  & $-$27.2(1)   & 1.05(11)   & 85  $\pm$ 14  & 119 $\pm$ 19\\
RT\,911125 & 8.4 & BnA  &  120 & 40.019(5)  & $-$27.5(1)   & 0.86(9)  & 58  $\pm$ 10  & 80  $\pm$ 13\\
RT\,900914 & 8.4 & B    &  70  & 39.951(7)  & $-$29.8(2)   & 2.12(13)   & 117 $\pm$ 13  & 172 $\pm$ 20\\
RT\,870814 & 5   & A    &  120 & 37.544(1)  & $-$22.22(3)  & 33.54(2) & 91  $\pm$ 5   & 75  $\pm$ 4 \\
RT\,871026 & 5   & BnA  &  70  & 39.996(4)  & $-$28.00(6)  & 0.87(5)  & 40  $\pm$ 3   & 54  $\pm$ 5 \\
RT\,871030 & 5   & BnA  &  120 & 39.994(4)    & $-$27.63(5)  & 1.04(5)  & 177 $\pm$ 13  & 239 $\pm$ 17

\enddata
\tablecomments{The top block contains the candidate events that thoroughly passed all tests and had smooth light curves in the event region throughout the observation (Level 2 candidates). The second block contains candidate events that have passed all tests (the sixth column in Table~\ref{tab:Archivaldata}), but did not have smooth light curves over their respective observation (Level 1 candidates). The bottom block contains the potential events from A, BnA, and B configurations that either barely failed a test or did not receive the full array of tests (ie. had only one polarization; Level 0 candidates). 
The transients are named for the UTC date on which they occurred in the form RT\,YYMMDD.  The duration is the time from the appearance of the transient signal to the disappearance, regardless of whether the emission was above the cutoff threshold at all times.  The positions ($\Delta$RA, $\Delta$DEC) are measured as offsets from ${\rm 17^h45^m00^s}$ and $-29^\circ00'00''$.  The distance ($\delta\theta$) is measured from \sgra\, at RA = ${\rm 17^h45^m40.06^s}$, DEC = $-29^{\circ}00\arcmin28\farcs20$.  The integrated and peak flux densities in the last two columns are reported for the maximum of the transient events.
Positions and integrated flux densities are derived using the \texttt{imfit} routine in CASA, 
which fits 2D gaussians to input sources assuming a gaussian noise background. The local rms around each transient is propagated to account for uncertainties from the undulating background. Because the fitting routine assumes uncorrelated noise, the uncertainties are almost certainly underestimated.\newline
$^\dagger$ The integrated flux is discrepant from the peak flux density due to insufficient model subtraction and sparse u-v coverage. We assume the peak flux density when calculating rates and speculating on source classes.
}
\label{tab:Transients}

\end{deluxetable*}

In searching for transient events, we have chosen to model the background flux for 
each observation instead of creating a global model from all the available data
in the archive.  A global model would provide a higher fidelity representation of the 
flux in the region (as a result of much better u-v coverage) and would be sensitive to 
the long-duration transients ($\Delta t_{\rm dur} \gtrsim T_{\rm obs}$) 
that get missed when averaged into a single observation model. However, the residual 
images created from subtracting a global model could have errors in the 
cross-calibrations between different observations, introducing artifacts 
like constant offsets to the residual images. Our search is targeted at 
transients with durations less than a few minutes (which accounts for a small 
fraction of the integration time in almost all observations), and so we have chosen the
observation-based modeling approach to potentially avoid cross-calibration complications.

\subsubsection{Identifying Transient Events}

After removal of the full observation model, the model-subtracted visibility data 
is imaged without CLEANing in intervals of one, two, and six times the 
visibility sample time (typically 10\,s, but could also be 20\,s or 30\,s). 
For each observation of \sgra, the mean, root-mean-square (rms), maximum, 
and minimum values of each image are calculated as a function of time in order
to locate and track temporal variations in flux density. 
Excursions of typically 2 or 3 standard deviations in the maxima between consecutive 
images are noted as potential transients, which usually corresponded to peak excursions
of about seven times the image rms.  

The potential transients are then inspected by eye in the residual images to ensure that 
events are physically plausible and not spatially dispersed or moving in the image domain. 
Additionally, the five brightest pixels in all images are identified and small 
regions around these pixels are tracked over all scans in the observation.  
This process isolates potential candidates and excludes from 
further consideration regions that have a constant excess in flux density due 
to insufficient CLEANing of the full observation model. 
Since events have to occur with statistical significance at the same location in 
consecutive 10\,s images to resolve a rise and fall, this search is sensitive 
to transients with timescales down to about 30\,s. 

\begin{figure}[ht]
\centering
\includegraphics[scale = 0.45]{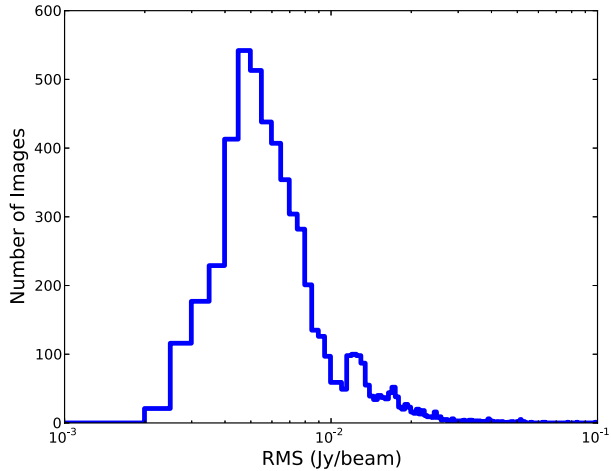}
\caption{ \footnotesize A histogram of the rms values for 10\,s residual images from all 8 GHz observations of the Galactic center in A configuration. The histogram is roughly Gaussian with a peak rms of $5~{\rm mJy~beam}^{-1}$. The extended tail towards higher rms values represent datasets that were either miscalibrated or improperly flagged. The expected thermal noise rms for 8 GHz observations in A configuration is approximately 0.5 mJy. Due to sidelobe effects and unsubtracted structure, our distribution of rms values has a median of 6.4 mJy.}
\label{fig:RMSHist}
\end{figure}

As a result of the short integration times and complex structure present in the 
Galactic center, the residual images produced from the model-subtracted 
visibility data are contaminated with large-scale ripples and sidelobes.  
In an attempt to avoid spurious detections, we set a fairly high threshold for  
identifying a potential transient candidate (typically six to seven times the image rms).  
The distribution of image rms values for all of the 10\,s residual images 
made from the 8~GHz A-configuration data is shown in Figure~\ref{fig:RMSHist} and a 
breakdown of values by frequency and configuration is presented in Table~\ref{tab:rms}.  
In addition to false-positives from noise, imaging surveys for radio transients 
must also guard against more subtle imaging artifacts.  
The careful reanalysis by \citet{fko12} of an archival VLA transient survey 
by \citet{bsb+07} found that many of the claimed detections were actually 
imaging artifacts created by insufficient CLEANing, imperfect calibration of the data, 
or other subtle antenna- or baseline-based errors.  
To filter out false-positives from such effects, we adopt a rigorous series 
of confirmation tests for each potential transient candidate. Candidates that 
failed at least one of these tests or did not receive the full suite of tests 
will be referred to as Level 0 candidates. Candidates that passed all tests will 
be referred to as Level 1 candidates, and candidates that are particularly 
promising will be referred to as Level 2 candidates.

\subsubsection{Event Confirmation}

After observations containing potential candidates events are identified, 
they are re-analyzed using subsets of the data in an attempt to reject 
false-positives caused by imaging artifacts or RFI.  A failure to detect the 
transient in the same location with comparable flux densities in all subsets 
will result in the rejection of the candidate event. Candidate events that 
passed all these tests (Level 1 and 2 events) are listed in the top two 
blocks of Table~\ref{tab:Transients}. In the first test, the full observation 
is split into two polarizations (right/left circular polarization) and separately 
run through the imaging and processing pipeline.  Though this test may reject real 
circularly polarized astrophysical sources, it will also effectively remove certain 
types of RFI. 

In the second test, the full observation is imaged separately in each of the two 
intermediate-frequency (IF) bands processed by the VLA.  For almost all of the observations
considered in this archival search, the difference in center frequencies between the two
IFs is ${\leq}100~\rm MHz$.  Thus, the detection of a transient event in one band but not
the other would indicate that the source is either a broadband source
with an extremely steep spectrum, a source with narrow line-like emission, or 
(most probably) narrow-band RFI.  

In the final test, the data are split into different subarrays and re-imaged to test
for corruption by a bad antenna or baseline.  The simplest way to isolate a single
bad antenna is to create two disjoint sets of antennas with similar u-v coverage.  
However, the reduction of sensitivity in this case is enough to make detection
of any of the transient events difficult.  To reclaim some of the lost sensitivity,
the antennas are instead split into three groups of 18 antennas where each group
has some overlap with the others, but each antenna is absent from one of the groups.
Detecting the candidate in all three antenna groups means that the transient could 
not be the result of a single bad antenna or a single bad baseline.  

The results of the confirmation tests are summarized in Table~\ref{tab:Transients}.
Level 1 and 2 candidates are listed in the first two
blocks of Table~\ref{tab:Transients}. While Level 1 candidates passed all 
our confirmation tests, their light-curves behaved erratically over the observation, 
suggesting they may potentially be a processing artifact. The first block contains 
the two highest quality candidates (Level 2) that passed all the tests and had a 
smooth, well-behaved lightcurve. The third block of Table~\ref{tab:Transients}
contains Level 0 candidates in the most extended array configurations that either
could not have the full suite of tests run (e.g., only one polarization), or 
marginally failed one of the tests.  

\subsubsection{False Positive Rate}
As one final check on our candidate events, we can estimate 
the false positive rate expected for the transient search. For each configuration
and frequency, the number of independent samples in an image is roughly the 
number of synthesized beams in that image, 
$N_{\rm b} \sim (\theta_{\rm FOV} / \theta_{\rm beam})^2$.  
If snapshot images are made with duration $\Delta t$, then the total number of 
independent samples in a data set with $T_{\rm obs}$ of total observing time 
is $N = N_{\rm b} \times (T_{\rm obs} / \Delta t) \approx 3.7 \times 10^9$.
We note that this rough estimate does not account for sidelobes 
and other unmodeled structure that are correlated across residual images.
Phase errors from atmospheric delays and instrumental errors could also
potentially lead to enhancements of signal.
We do find that the distribution of changes in pixel values
from one residual image to the next is roughly Gaussian, with systematic
sidelobe structure varying on longer timescales than the transients we detect.  
We search for excursions that are localized in time and position on the sky, 
and note when transients are in regions of high sidelobe activity to minimize the effect
of unmodeled structure.

Assuming the noise in each of the residual images 
is roughly Gaussian, with the parameters of the Gaussian varying between images,
we expect $N_{\rm FP}(S \geq 5.8\,\sigma) = 12$ 
false positives with significance at or above the $5.8\,\sigma$ level set by the 
weakest event detected in our search. This rough estimate also does not include the 
requirement that events occur in a single location in consecutive snapshots (which would dramatically
reduce the false positive rate), nor does it explicitly account for the effects
of unmodeled structure (which artificially increase the $\sigma$ parameter in each of our
residual images). However, it provides a useful
comparison to the 23 Level~0, Level~1, and Level~2 candidates produced by the detection 
pipeline.  Since only two of these events are classified as Level~2 candidates,
it is likely that the false positives are largely accounted for by 
our confirmation tests.

\section{Results}
\label{sec:results}

\begin{figure*}
\centering
\includegraphics[width = 0.49\textwidth]{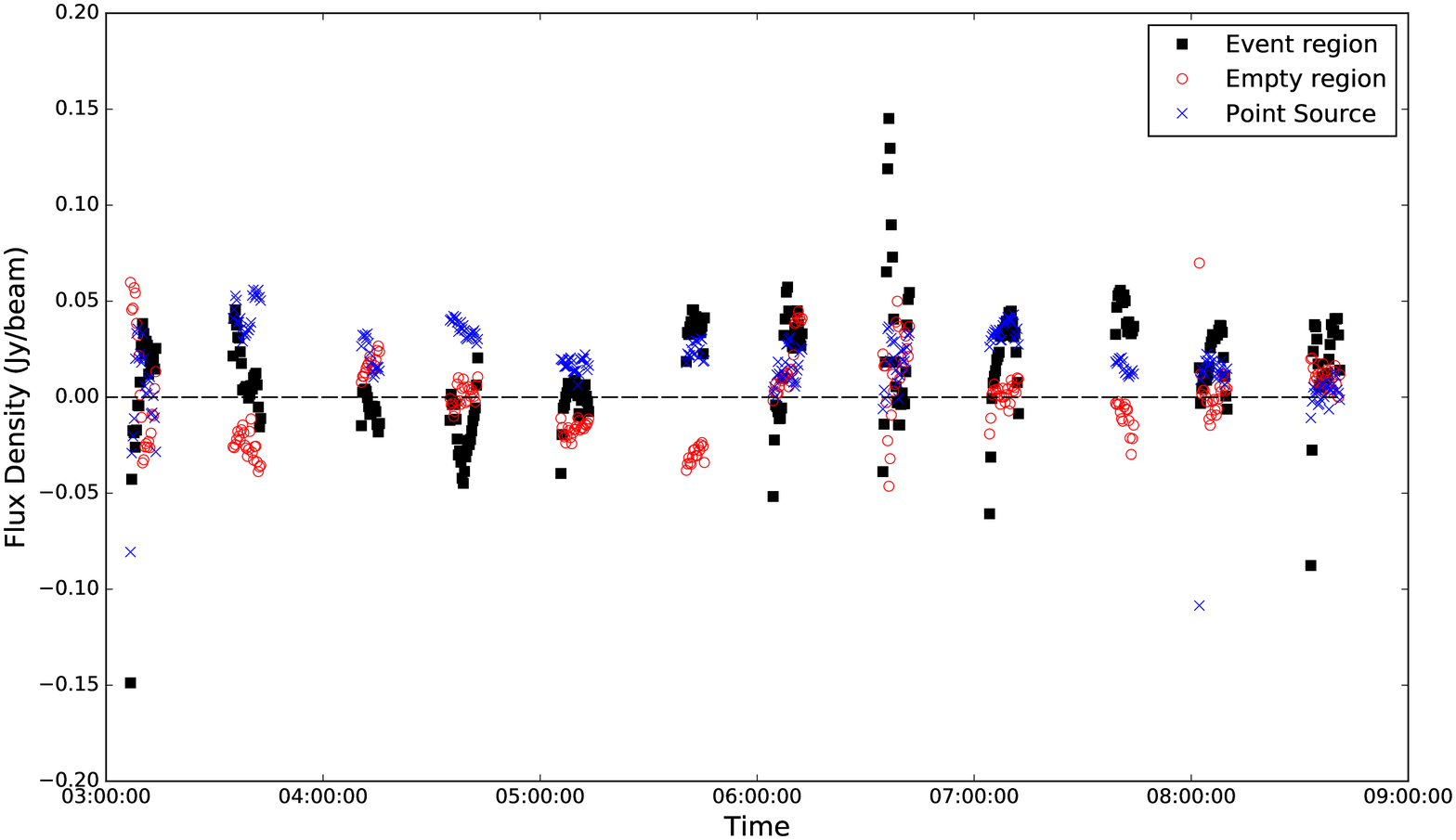}
\includegraphics[width = 0.49\textwidth]{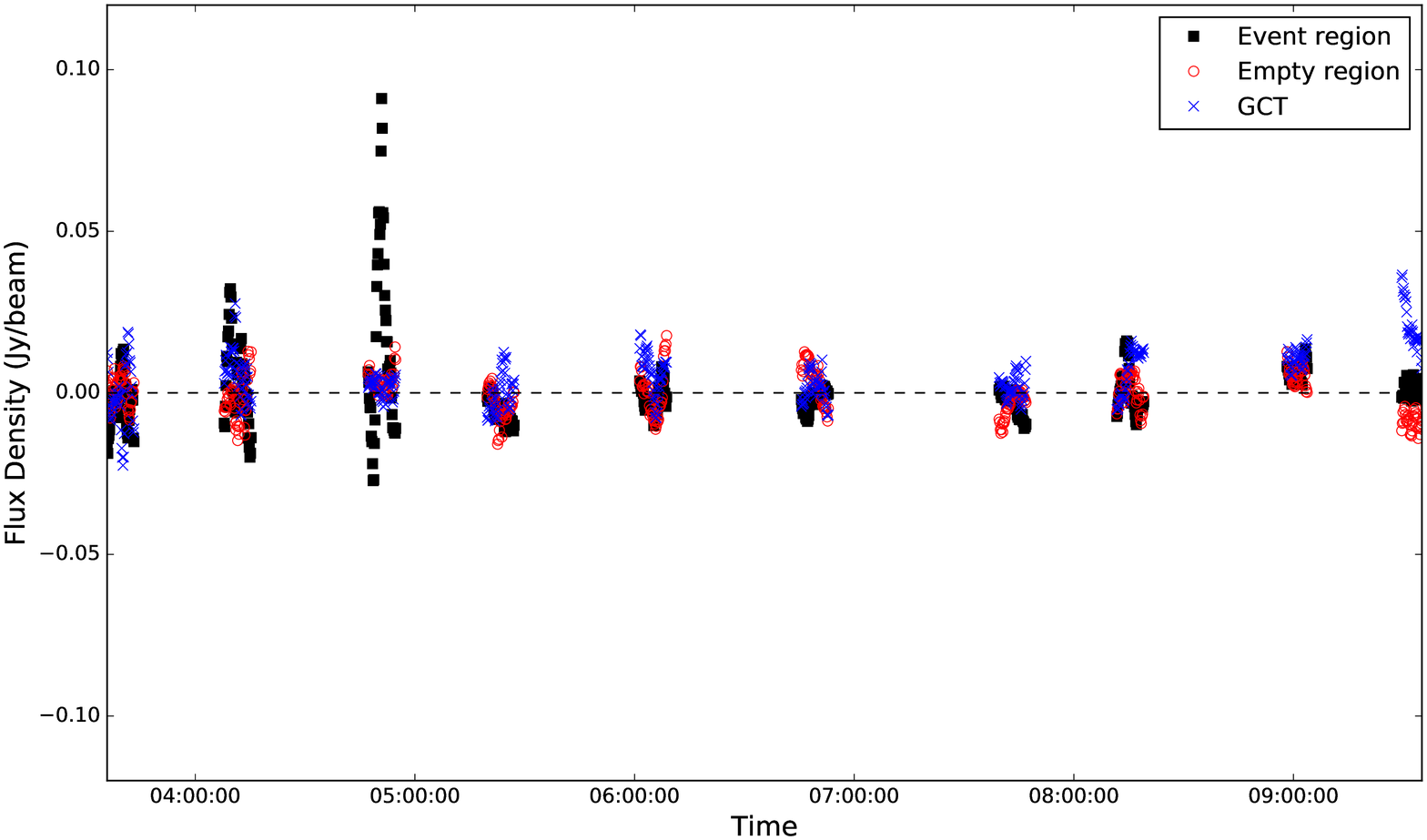}
\caption{\footnotesize Plots of the peak flux density of the burst vs. time for \Eone\ in 20\,s cadences (left) and \Etwo\ in 10\,s cadences (right) in model-subtracted images. The black squares correspond to the flux density of the event over the duration of observation, red points are the flux density of an empty region during the event, and blue points are the flux density a reference source in each field, plotted for comparison. Gaps in the light curves are largely due to time on other fields}. All times and dates are in UTC.
\label{fig:2events}
\end{figure*}

\begin{figure*}
\centering
\includegraphics[width=0.49\textwidth]{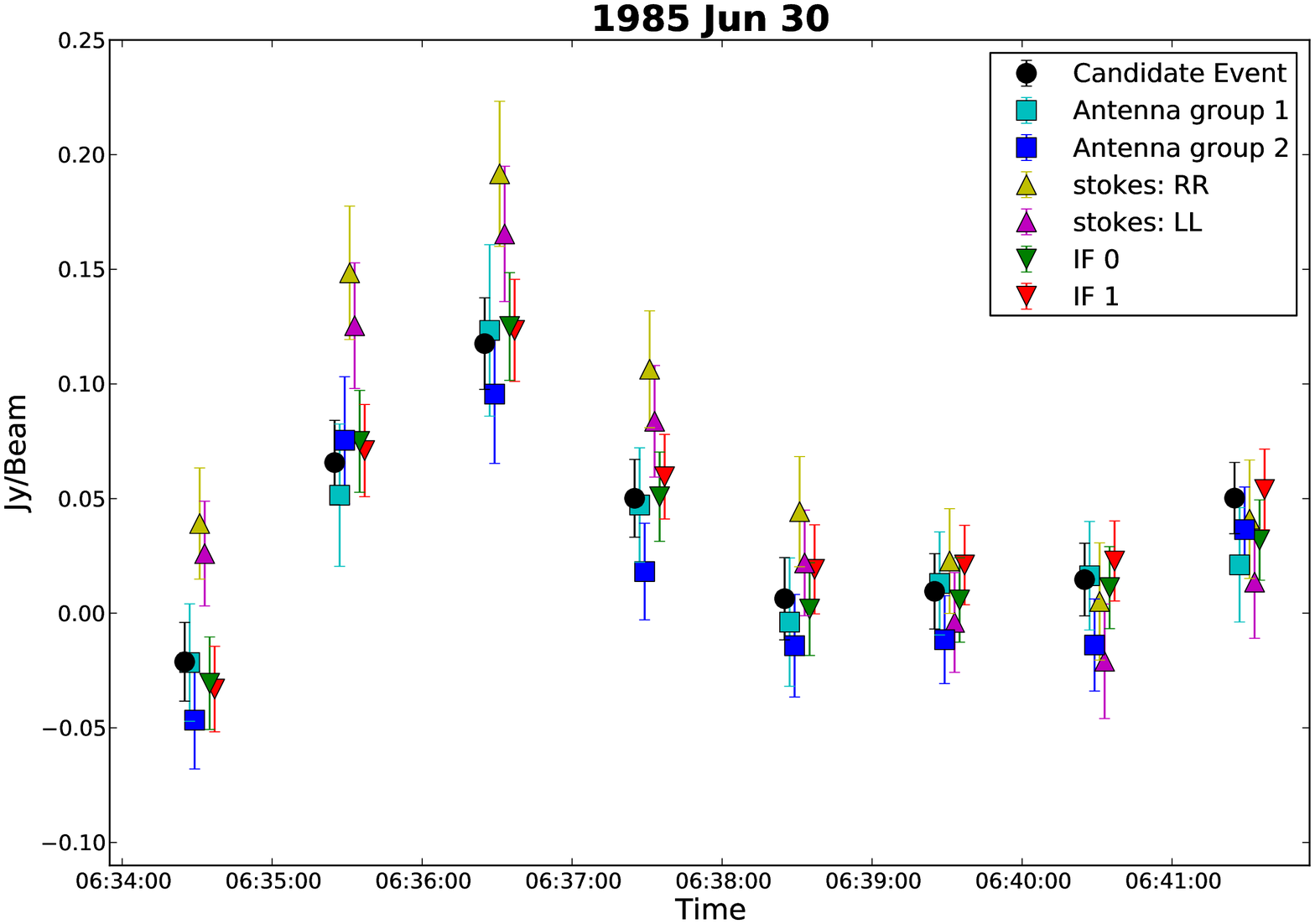}
\includegraphics[width=0.49\textwidth]{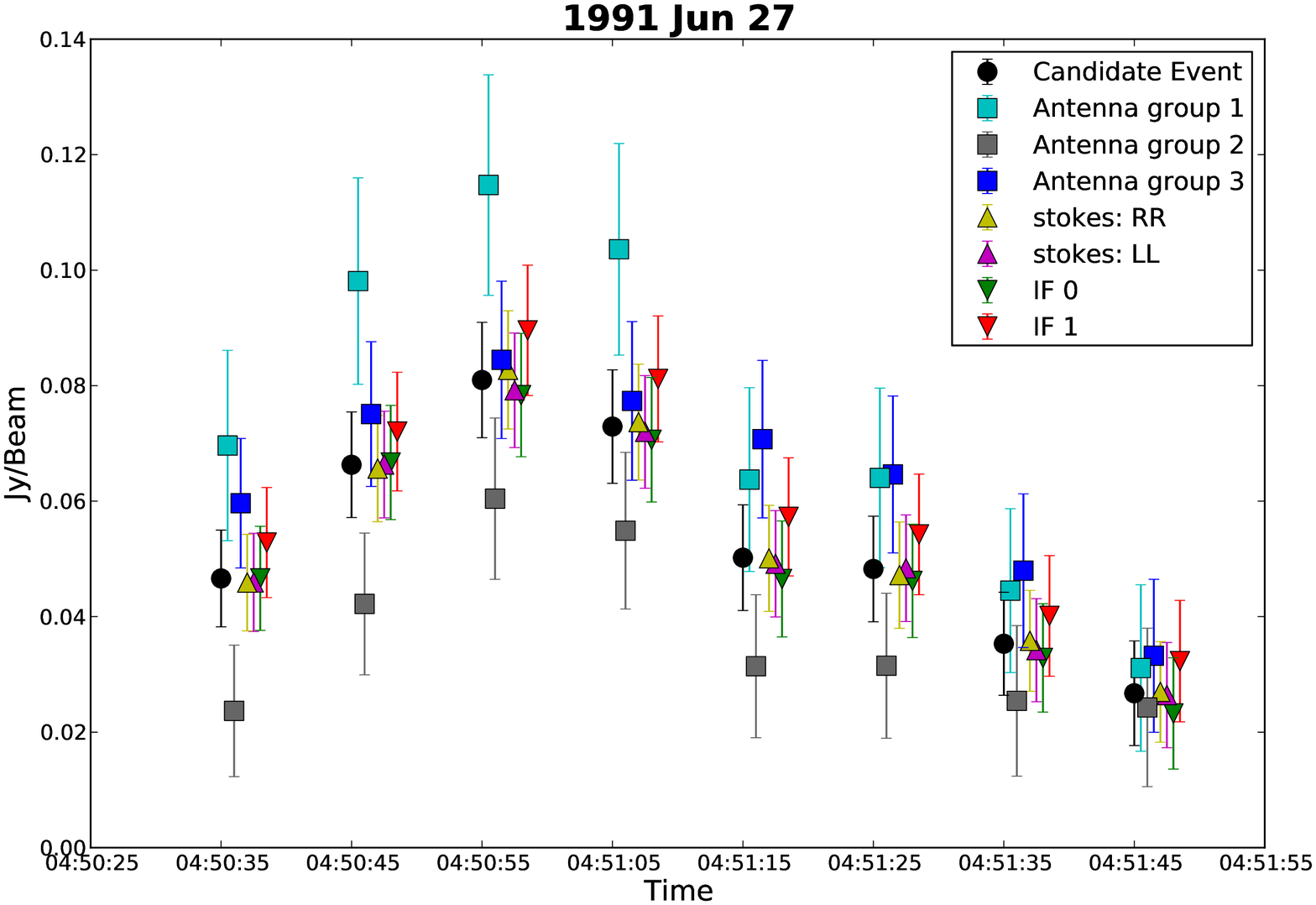}
\caption{{\sc Left}: light curves for \Eone\ in 60\,s residual images for different groupings of data:  data split by spectral windows, data split by stokes parameters (LL, RR), and finally, data split between two antenna groups. The light curves labeled by each symbol are offset by 5s of each other for clarity.
{\sc Right}: light curves for \Etwo\ in 10\,s residual images for different groupings of data:  data split by spectral windows, data split by stokes parameters (LL, RR), and finally, data split between three antenna groups such that each antenna is missing in at least 1 group. The light curves labeled by each symbol are offset by 0.5\,s for clarity. Error bars correspond to the rms of the residual image. All times and dates are in UTC.}
\label{fig:splits}
\end{figure*}


\begin{figure*}
\centering
\includegraphics[width = 0.49\textwidth]{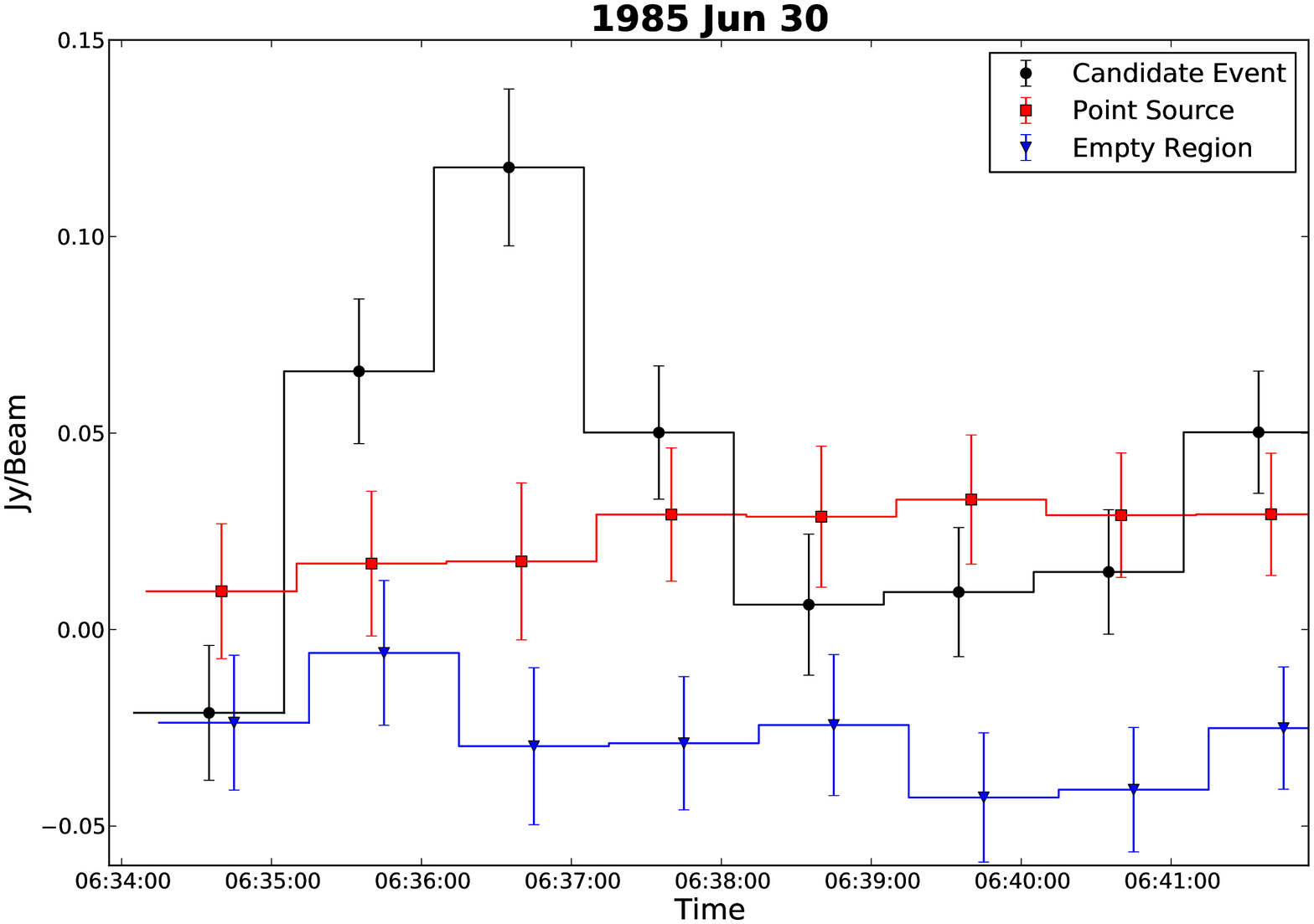}
\includegraphics[width = 0.49\textwidth]{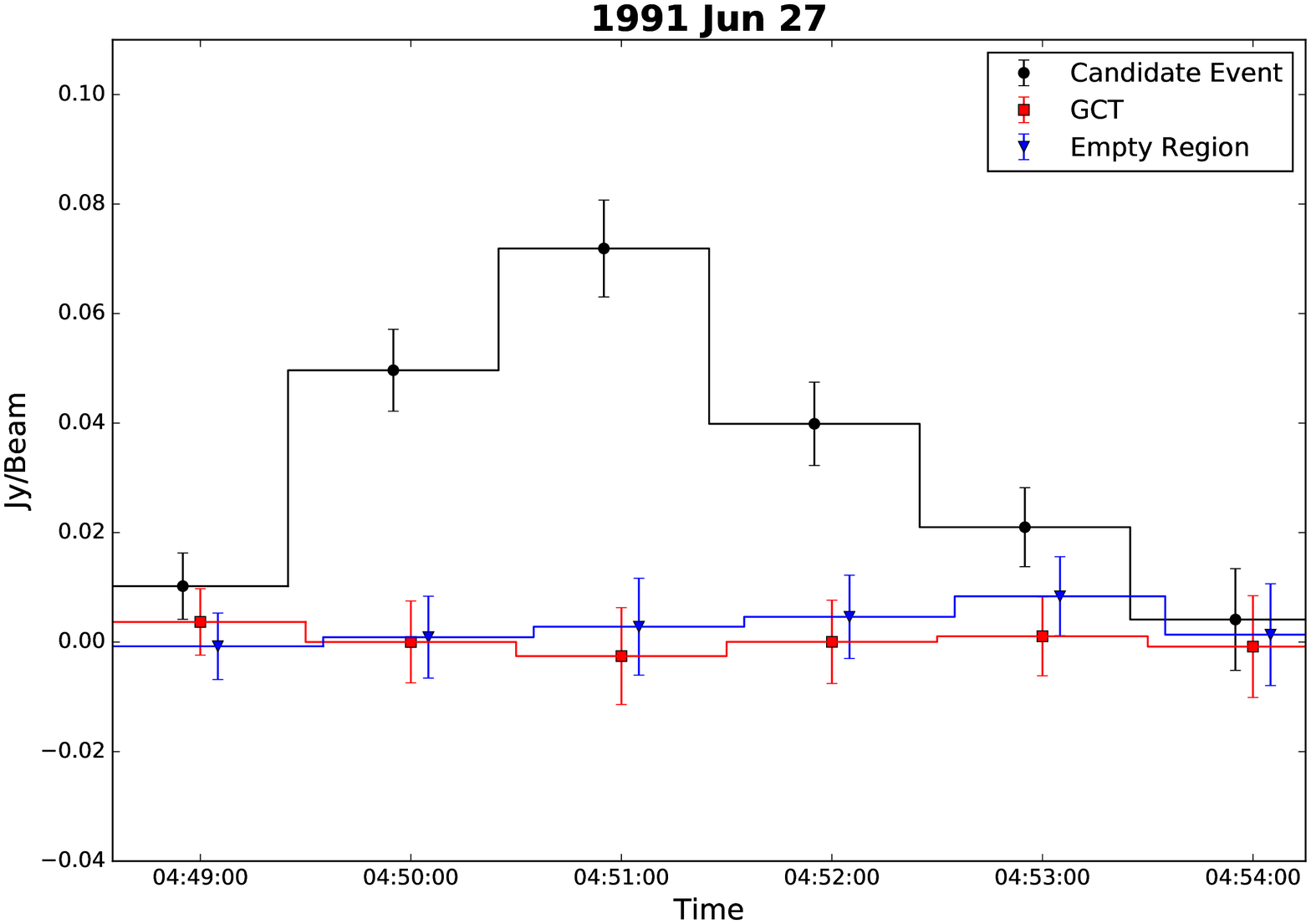}
\caption{{\sc Left:} light curves of \Eone, a compact source, and an empty region in the \sgra\, field from 60\,s residual images. 
{\sc Right:} light curves of \Etwo, the GCT, and an empty region in the \sgra\, field from 60\,s residual images. Both bursts occur with no corresponding excursions in the compact source for \Eone, the GCT for \Etwo, or the empty region. Error bars correspond to the rms of the residual image.
All times and dates are in UTC.}

\label{fig:diffsources}
\end{figure*}

\subsection{Candidate Events}
\label{ssec:events}

Of the 23 candidate transient events identified in our processing and 
detection pipeline, only eight passed all of the confirmation tests designed
to filter out RFI and imaging artifacts. The locations of these candidates
relative to \sgra\ are displayed in Figure~\ref{fig:events}. Of the 
eight Level~1 candidates, 
only two are deemed to be Level 2 candidates based on the smoothness of their 
lightcurves (Figure~\ref{fig:2events}). We note that while many of our candidate events
appear to be resolved, this is mainly due to excess flux from 
improper model subtraction, noise, and PSF-related effects (Figures~\ref{fig:psfAY8} and~\ref{fig:psfAZ52}).
In this section, we present details on these two events,
and discuss efforts to find counterparts at other wavelengths.

\subsubsection{\Eone}
\Eone\ was detected in project AY8, observed on 1985 June 30 at 5~GHz. The
event lasted roughly 120\,s, was located $4\farcs6 \pm 1\farcs9$
north of \sgra, and had a maximum peak flux density of $149 \pm 26~{\rm mJy\,beam^{-1}}$.

We split the data from project AY8 into different
subgroups (by frequency, sub-array, and polarization) and found a
consistent transient signal in each subgroup, as shown in
Figure~\ref{fig:splits}. Further, we constructed light curves of an
empty sky region anti-symmetric to the event with respect to the 
pointing center, 
as well as a point-like test source composed of the two objects
1LC 359.985+0.027 and 2LC 359.985+0.027 (RA${\rm = 17^h45^m28.66^s}$ 
Dec = --$28^{\circ}56\arcmin03\farcs943$)
in the same field \citep{lc+98}, in case the entire residual image had an artificially elevated flux for a brief
period of time. As shown in Figure~\ref{fig:diffsources}, only the
event region showed a rise in flux density. 
Our measured peak flux density of 143 mJy for the point-like test source is in agreement with the 
5 GHz peak flux density reported for the object by \citet{bwh+94}.
Due to the proximity of \Eone\, to \sgra, we are unable to meaningfully constrain its quiescent emission 
even with our deepest images of the Galactic center.
We find no matches to known radio sources in the Master Radio Catalog from the High Energy Astrophysics 
Science Archive Research Center (HEASARC) archives.

\begin{figure*}
\centering
\includegraphics[width = 0.98\textwidth]{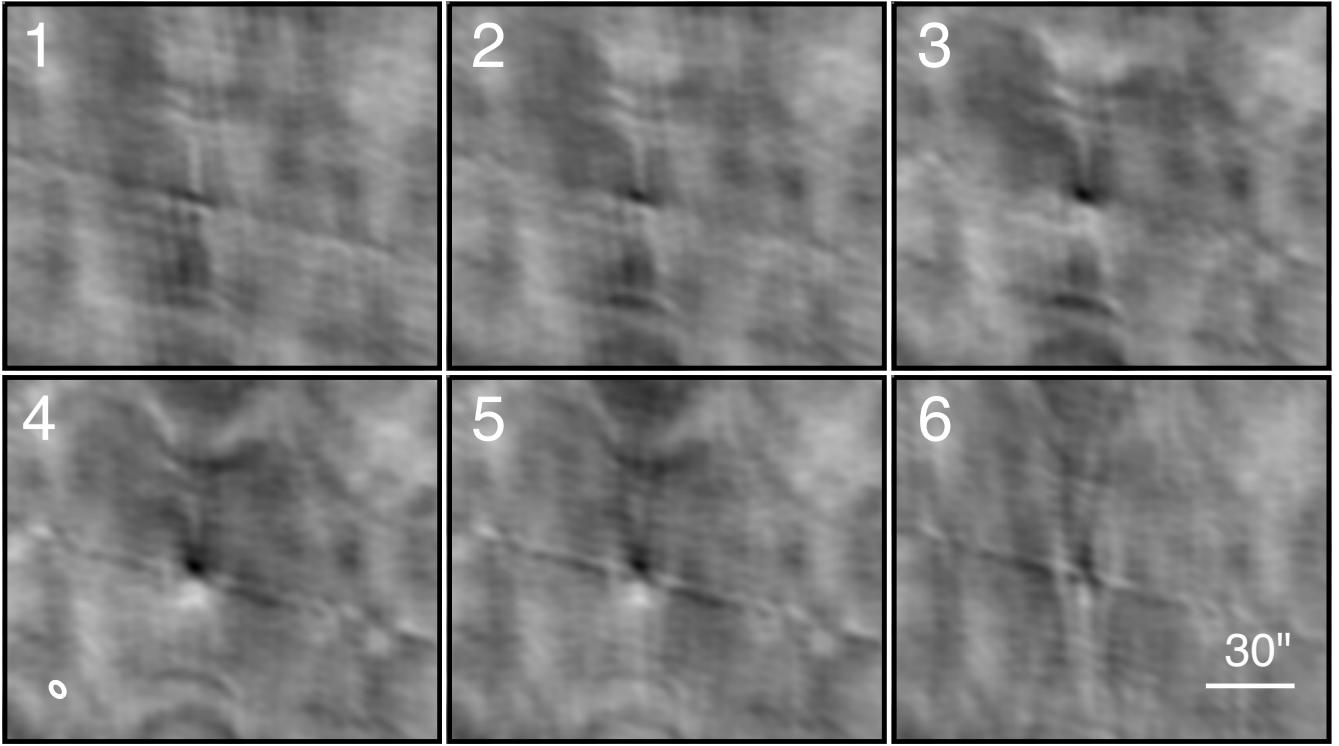}
\caption{\footnotesize 20\,s residual images corresponding to \Eone, with the greyscale ranging from $-0.15~{\rm mJy~beam}^{-1}$ to $0.15~{\rm mJy~beam}^{-1}$ and linear scaling. The images span UTC 6:35:25 to 6:37:25 on June 30, 1985. Each snapshot image is produced with u-v data from which a model for \sgra\, has already been subtracted. The region of negative flux below \Eone~has a peak amplitude amplitude $\sim60\%$
as large as the peak amplitude of \Eone. A uniform amplitude scale has been applied to each image, and the beam is shown in the lower left of the figure. North is up and East is left. Each panel of this figure is available as the Data behind the Figure. }
\label{fig:AY8}
\end{figure*}

Since the residual visibilities are imaged on short timescales,
each image has only snapshot u-v coverage, and incompletely
subtracted structure around \sgra\ produces large sidelobes. As such,
there is a risk that in spite of our tests, \Eone\ is an
artifact. Furthermore, \Eone\ appears in a region of high residual 
image flux due to poor model subtraction (Figure~\ref{fig:AY8}).
In Figure~\ref{fig:psfAY8}, we verify that the dominant side lobe pattern at the 
location of the transient event is traced by the PSF.
Thus, the sidelobe pattern in the residual image was probably produced
by the transient event itself, although we cannot definitively rule
out an imaging artifact. 
We also generated 20\,s images of the two minutes of transient activity without 
prior model-subtraction and detected a flux increase comparable to the flux of \Eone.
Finally, we conducted an independent re-analysis
of the data using the 
AIPS\footnote{\texttt{http://www.aips.nrao.edu}}
software package and detected the transient at the same location and time in project AY8.

\subsubsection{\Etwo}
\begin{figure*}
\centering
\includegraphics[width = 0.98\textwidth]{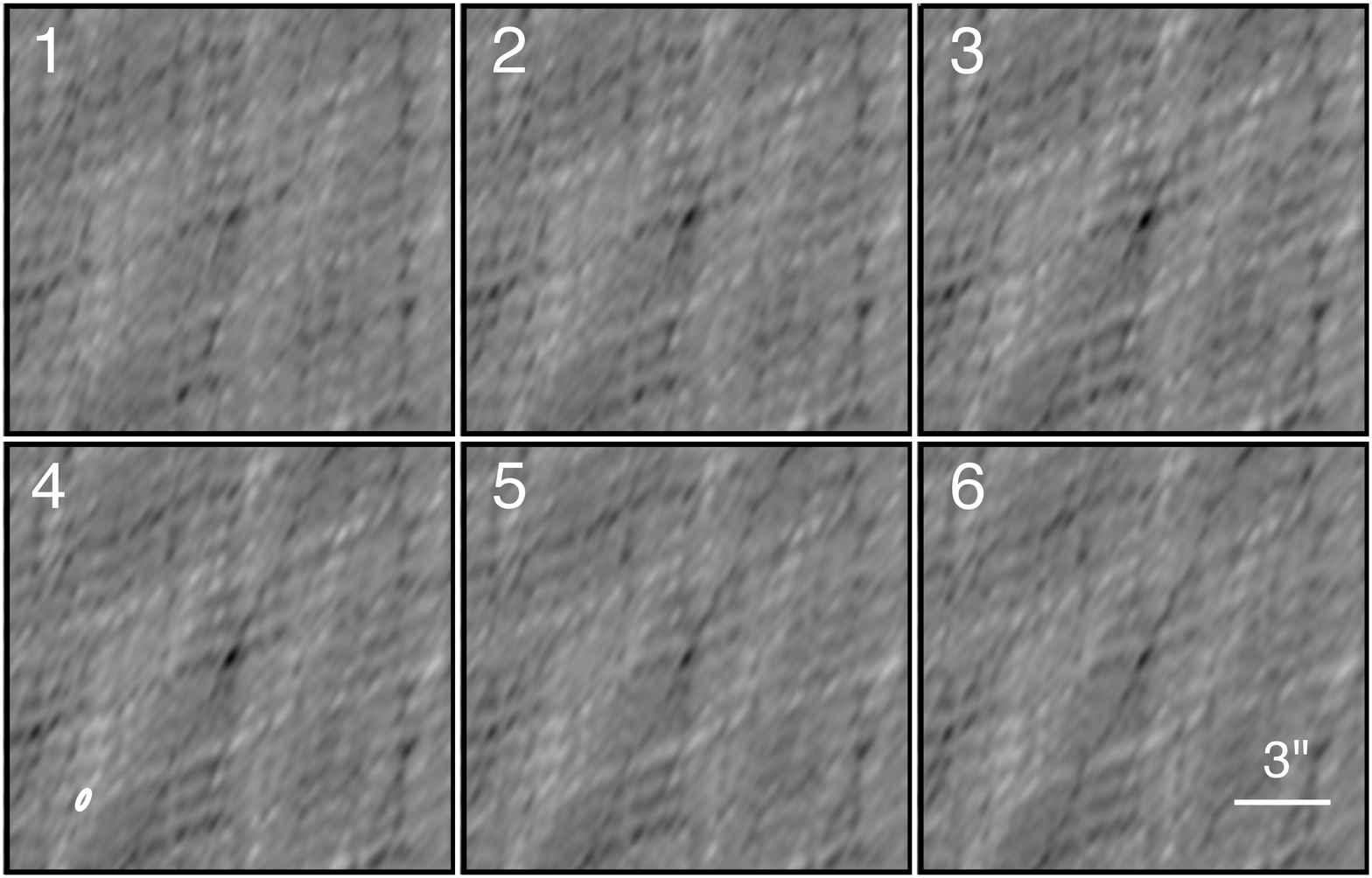}
\caption{\footnotesize 10\,s residual images corresponding to \Etwo, with the greyscale ranging from $-0.095~{\rm mJy\,beam}^{-1}$ to $+0.095~{\rm mJy\,beam}^{-1}$ and linear scaling. The images span UTC 4:50:35 to 4:51:35 on June 27, 1991.
See caption for Figure~\ref{fig:AY8}. Each panel of this figure, after a primary beam correction,
 is available as the Data behind the Figure. }
\label{fig:AZ52}
\end{figure*}

\begin{figure*}[ht!]
\centering
\begin{minipage}{.26\textwidth}
\centerline{\includegraphics[width = \textwidth]{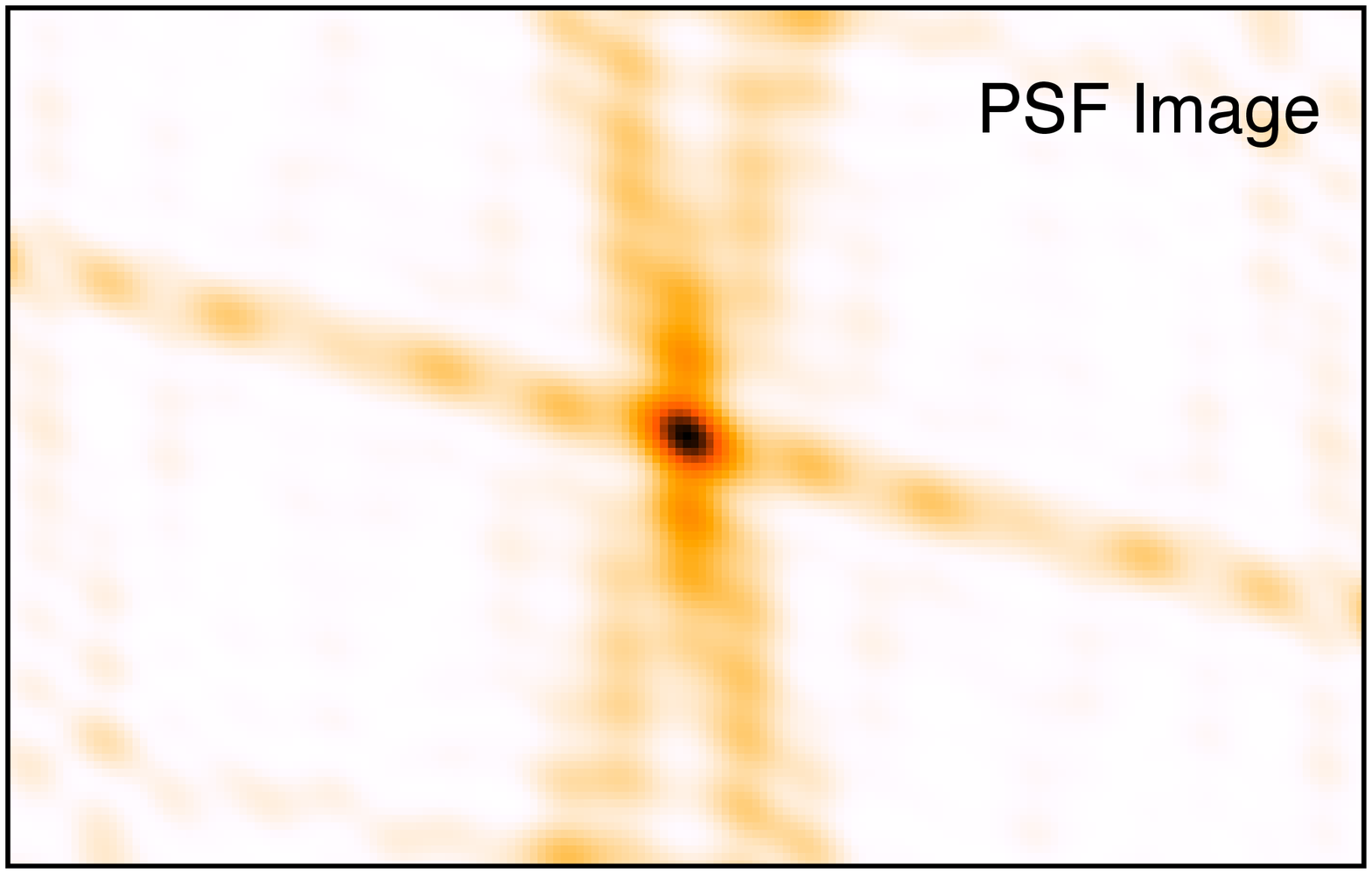}}
\end{minipage}
\begin{minipage}{.26\textwidth}
\centerline{\includegraphics[width = \textwidth]{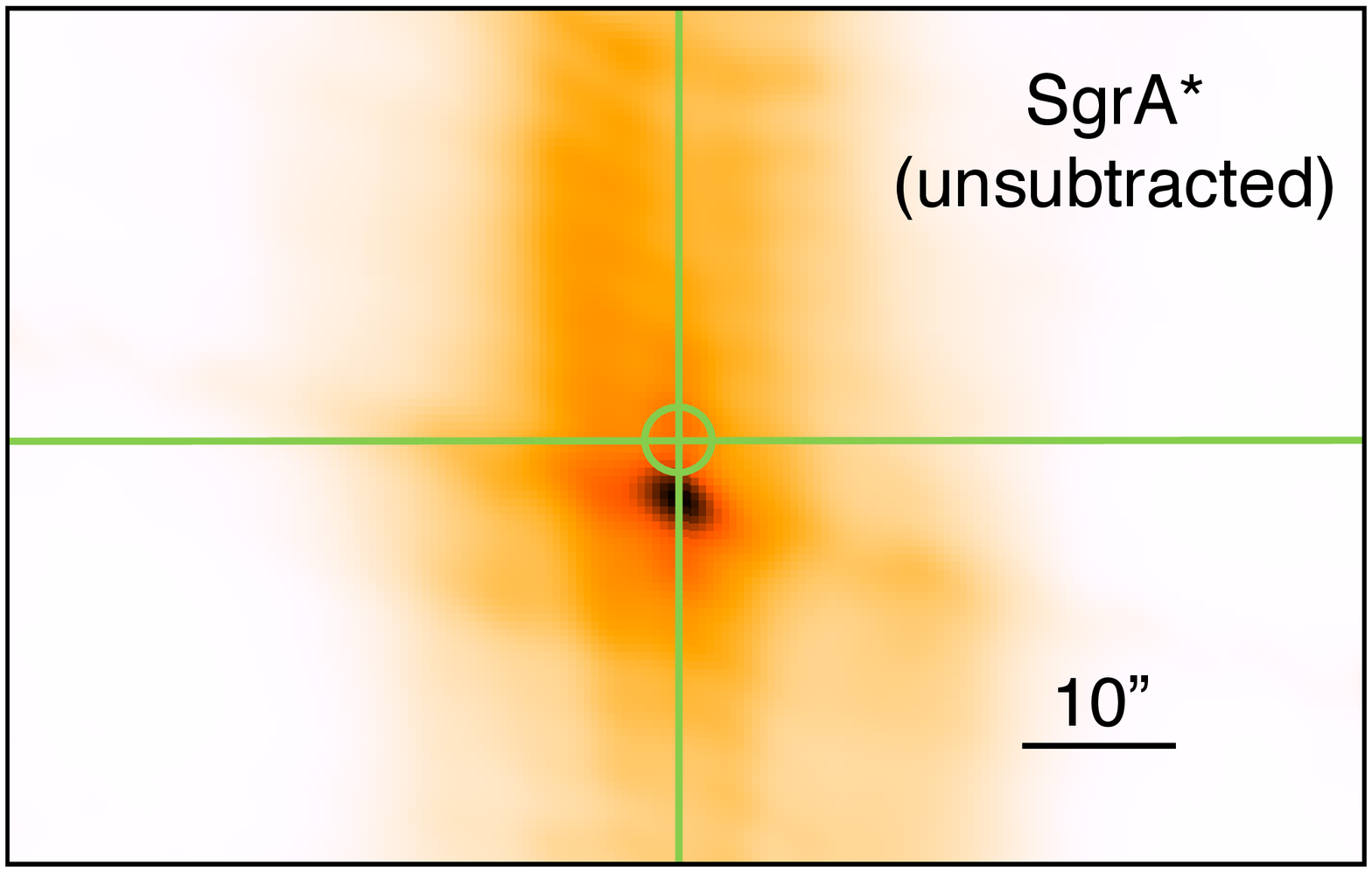}}
\end{minipage}
\begin{minipage}{.26\textwidth}
\centerline{\includegraphics[width = \textwidth]{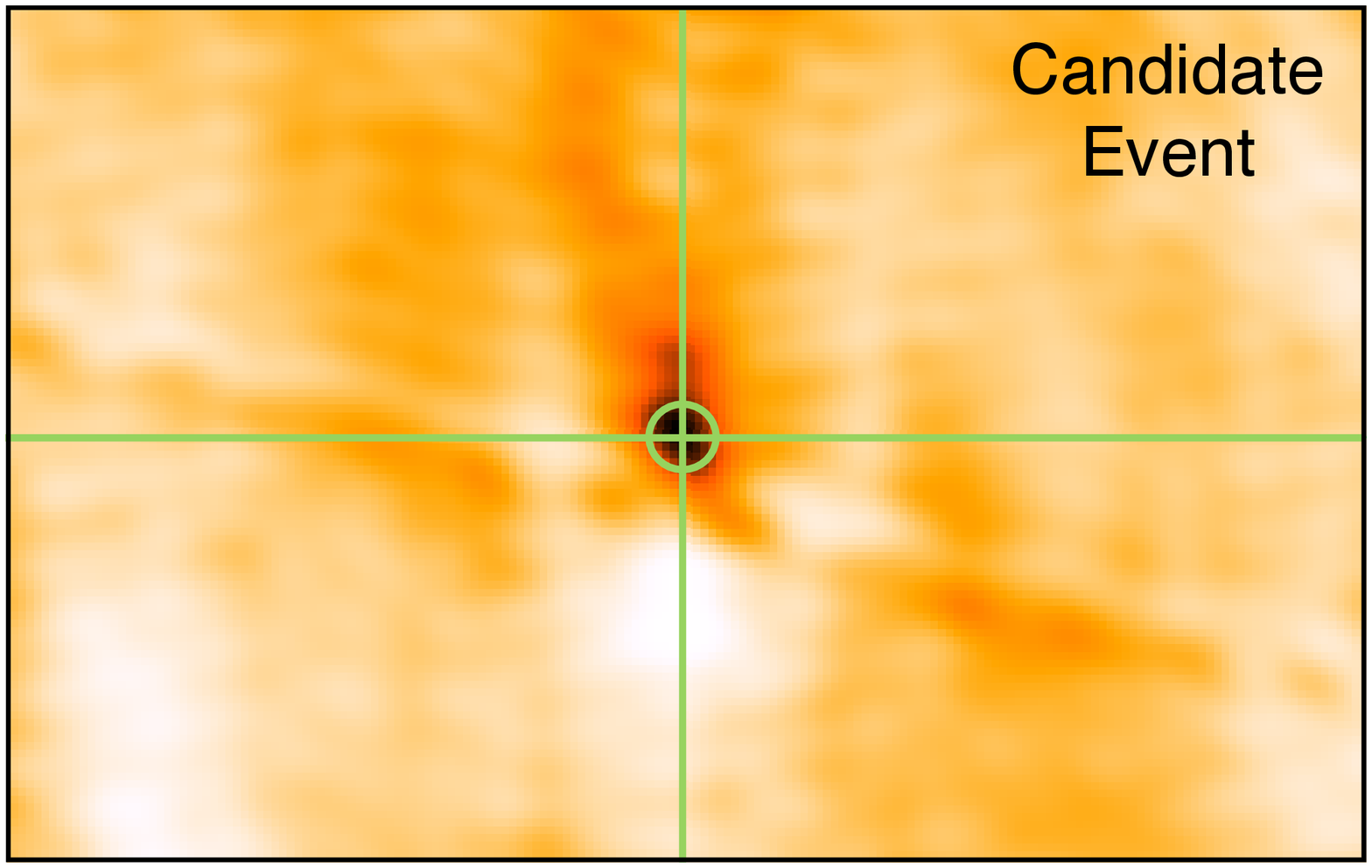}}
\end{minipage}
\caption{{\sc Left}: The point spread function {\sc Center}: An image of \sgra\, centered on \Eone in the 60\,s time bin corresponding to the
 peak of \Eone. {\sc Right}: 60\,s image with \sgra\, model subtraction, showing \Eone. The vertical side lobe pattern at the location of \Eone\ is traced by the PSF.}
\label{fig:psfAY8}
\end{figure*}

\begin{figure*}[ht!]
\centering
\begin{minipage}{.26\textwidth}
\centerline{\includegraphics[width = \textwidth]{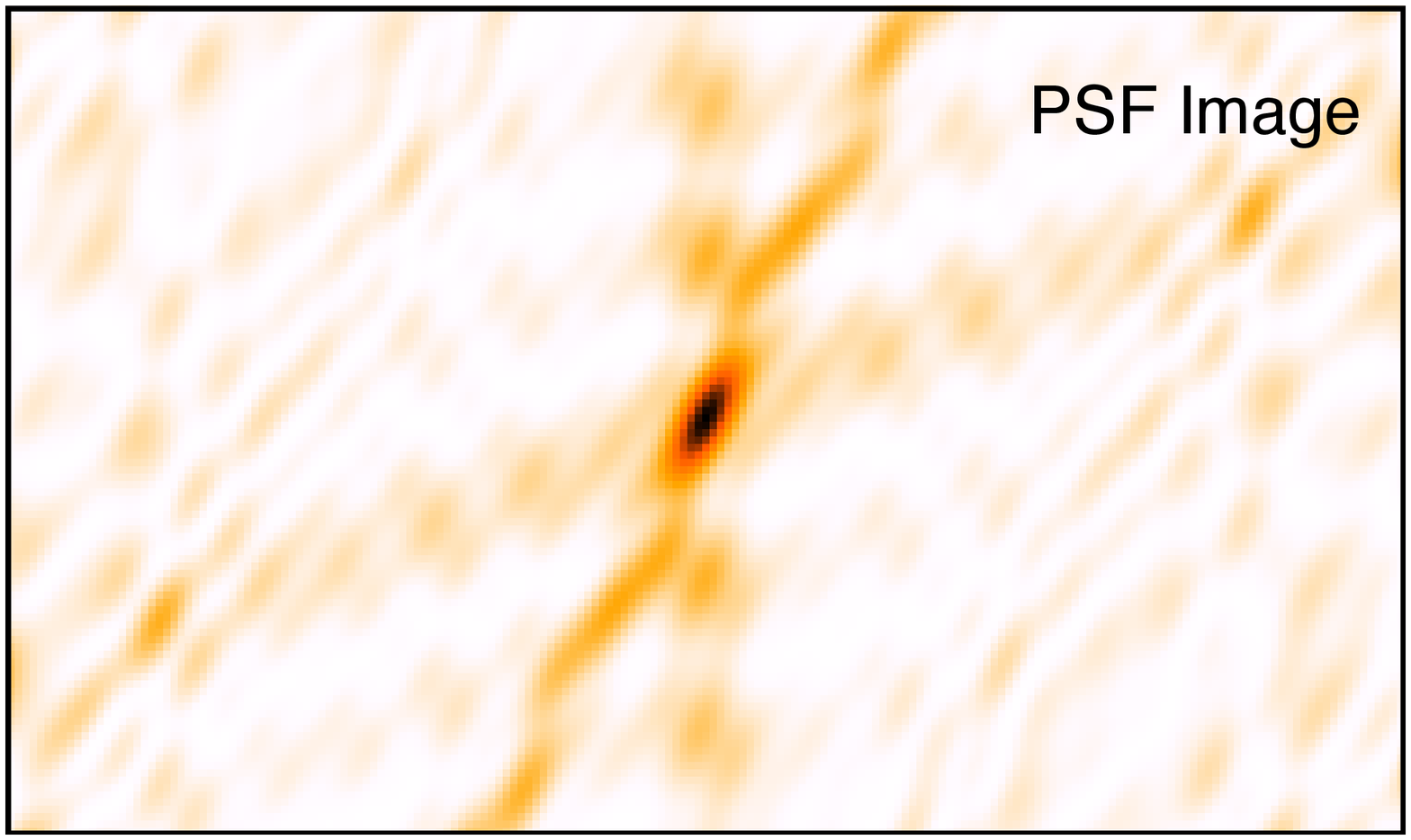}}
\end{minipage}
\begin{minipage}{.26\textwidth}
\centerline{\includegraphics[width = \textwidth]{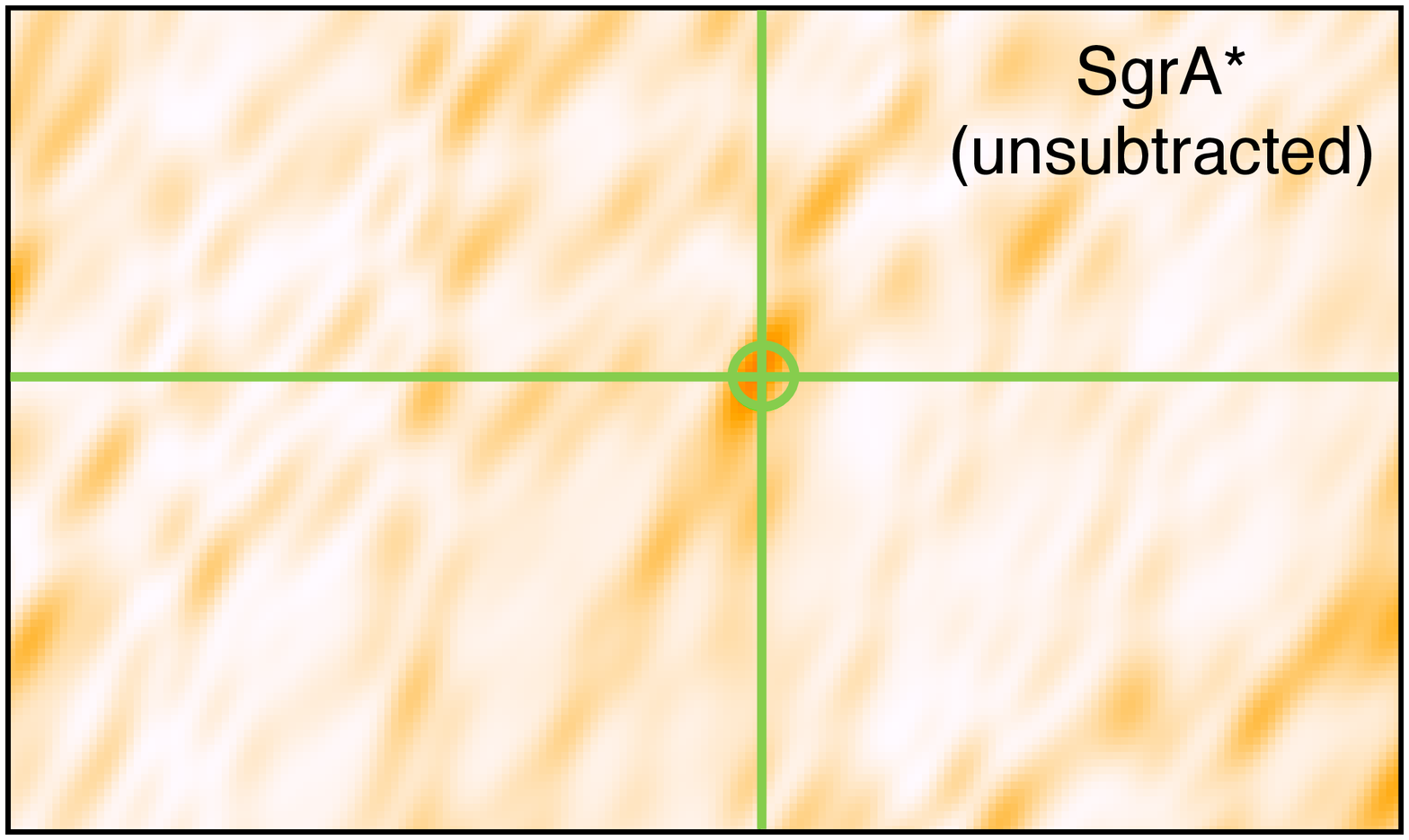}}
\end{minipage}
\begin{minipage}{.26\textwidth}
\centerline{\includegraphics[width = \textwidth]{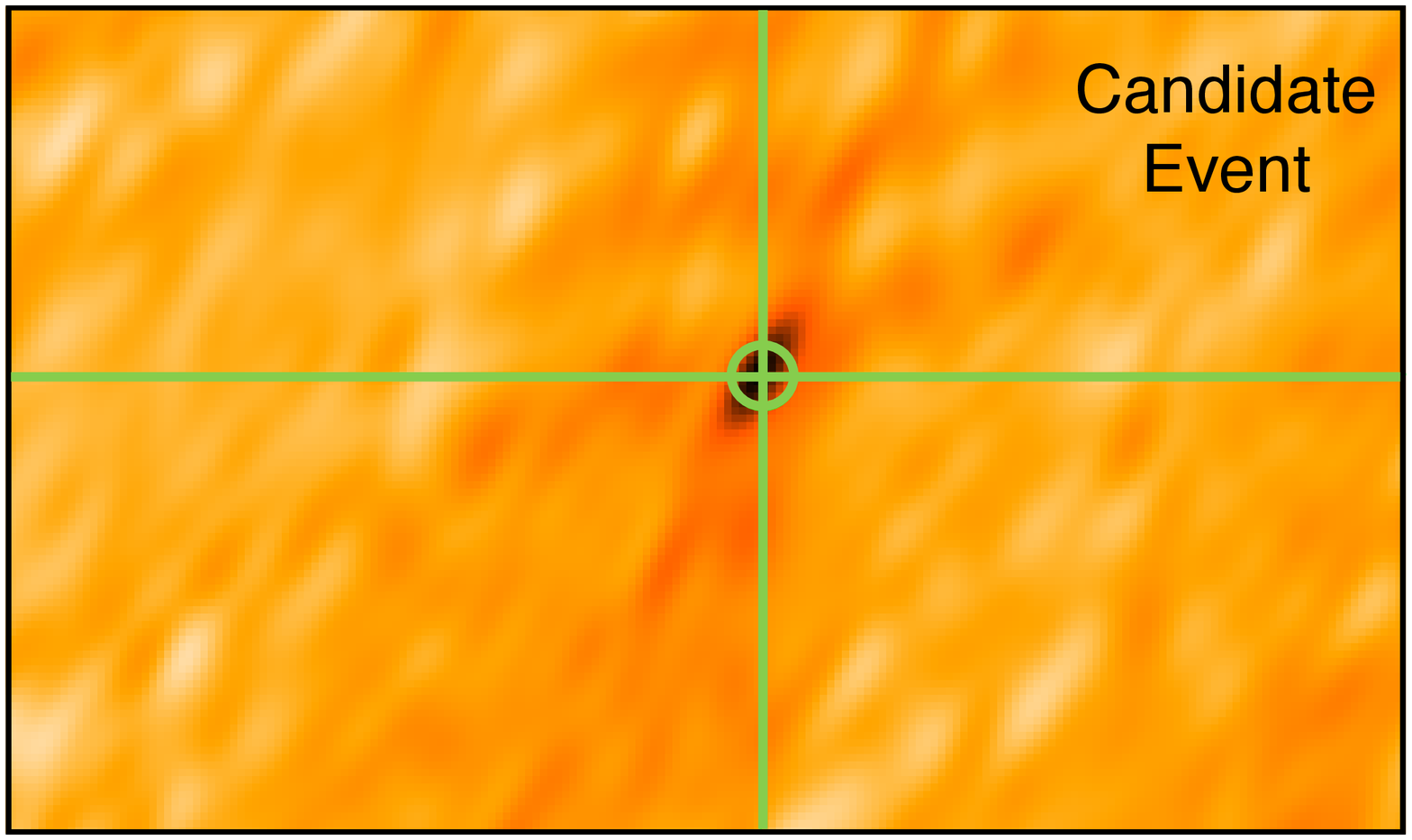}}
\end{minipage}
\caption{{\sc Left}: The point spread function {\sc Center}: An image of \sgra\, centered on \Etwo in the 10\,s time bin corresponding to the
 peak of \Etwo. {\sc Right}: 10\,s image with \sgra\, model subtraction, showing \Etwo. The side lobe pattern at the location of \Etwo\ is traced by the PSF.}
\label{fig:psfAZ52}
\end{figure*}

\Etwo\ (see Figure~\ref{fig:AZ52}) was detected in project AZ52, observed on 1991 June 27 at 8.4~GHz. It
lasted roughly 100\,s, was located roughly 20\arcsec\ northwest of
\sgra, and had an maximum peak flux of $100 \pm 11~{\rm mJy\,beam^{-1}}$.
We estimate that the effect of bandwidth smearing amounts to roughly a 10\% decrease
in peak flux density, based on the equations for bandwidth smearing in
\citet{bs+99}, for which our reported measurement is not corrected. We also find no matches to 
known radio sources in the Master Radio Catalog from the HEASARC
archives.

\Etwo\ passed all tests for validity and survived
each step of the additional analysis as described above for \Eone,
using the long-duration galactic center transient (hereafter GCT) detected by \citet{zrg+92} 
as a substitute for a constant point-like source
to construct additional light curves. 
The quiescent emission of the candidate event can be constrained to be $\lesssim 1$\,mJy/beam
based on the flux of the event region in the image of the full observation.
We find that this event appears to be
split into two statistically insignificant sources if imaged without self-calibration,
and this event is not detected after reducing the data with the AIPS software package.
However, we performed an additional independent re-analysis of 
the data set using the CASA software package and detected the transient at 
the same location and time as the pipeline.
For the above reasons, this event remains more suspect as an imaging artifact than \Eone.

\subsection{Counterparts at Other Wavelengths}
Given the dates and locations of the two best candidate events, we can look for 
potential counterparts at other wavelengths.  First, we check to see if
our events are consistent with any known transients in the Galactic center.  
None of the events are coincident with the GCT \citep{zrg+92}, 
the magnetar \gcmag, or any of the transient X-ray binaries 
discovered by \citet{mpb+05}.  See Figure~\ref{fig:events} for a visual
overview of the region.

Next, we compare our events against known X-ray and near-infrared (NIR) point sources.
The 2~Ms \emph{Chandra} point source catalog of \citet{mbb+09} contains
63 X-ray sources within 25\arcsec\ (1~pc at 8.5~kpc) of \sgra.  One of these sources,
CXOGC~174540.1$-$290025, is $2\farcs6$ from \Eone, falling just inside the synthesized
beam of that observation.  Interestingly, CXOGC~174540.1$-$290025 is characterized
by \citet{mbb+09} as having short-term variability.  However, given the number of sources, 
the probability of getting at least one source within the relatively large synthesized beam 
of \Eone\ is about $0.985$, so not much significance can be attached to this association.  
\Etwo, which is much better localized and further from \sgra\ than \Eone, has no associated 
X-ray point source.


Using the \citet{sme+09} catalog of the positions and proper motions of over 6000
stars within a projected distance of 1~pc from \sgra, counterparts to the transient
events can be sought in the NIR.  Given the high density of sources and the large
synthesized beam, any association with \Eone\ is difficult to determine.  However,
one of the ${\sim}100$ sources within the synthesized beam is the bright 
M1 supergiant GCIRS~7.  There is only one source within the beam of \Etwo.  Given
the size of the beam and the number of sources in the region, the probability of a chance
overlap is about $0.63$, so the association is again tenuous.

Finally, the dates and times of the candidate radio transients can be checked against
known gamma-ray burst (GRB) catalogues.  There are no reported GRBs coincident with 
\Eone, although at the time the only operational gamma-ray detector was the 
\emph{Pioneer Venus Orbiter} \citep{keg+80, efk+81}.  GRB~910627 occurred at 
UT~04:29:23 on 1991 Jun 27, about twenty minutes before \Etwo.  However, 
the position of GRB~910627 is constrained to an error box of about one degree centered 
on ($\alpha = 13^{\rm h}17^{\rm m}29^{\rm s}$, $\delta = 02^{\circ}30\arcmin27\arcsec$), 
which rules out any possible origin near the Galactic center \citep{hbb+00}.

The lack of clear counterparts to our transient events at other wavelengths 
is not surprising since these events occurred undetected 25 to 30 years ago 
(1985 and 1991). There were no targeted efforts at followup, and no 
contemporaneous programs with regular observations of the Galactic center. 
Repeating the present analysis on recent observations with good multiwavelength coverage 
\citep[e.g., the recent G2 campaign; see][]{ggf+13}
might permit the identification of counterparts to detected transient events.

\section{Occurence Rate of Transient Sources}
\label{sec:Rates}

Assuming our top two candidates are real astrophysical transients, we can 
estimate the rate of occurrence of similar events as a function of peak 
flux density.  We provide two different types of rates here: the 
transient rate ($\rho$, events per unit time and solid angle) and the 
Galactic center rate ($r$, events per unit time).  The transient rate is 
the standard rate that accounts for the variable sky coverage for each observation 
and can be easily compared against similar rates for other transient sources.  The 
Galactic center rate, however, is useful if the population of potential transient 
events is entirely contained within the solid angle of the smallest image used 
in this survey, as is likely if the sources are associated with the Galactic center.  

\begin{figure}[bt]
\centering
\centerline{\includegraphics[width = 0.48\textwidth]{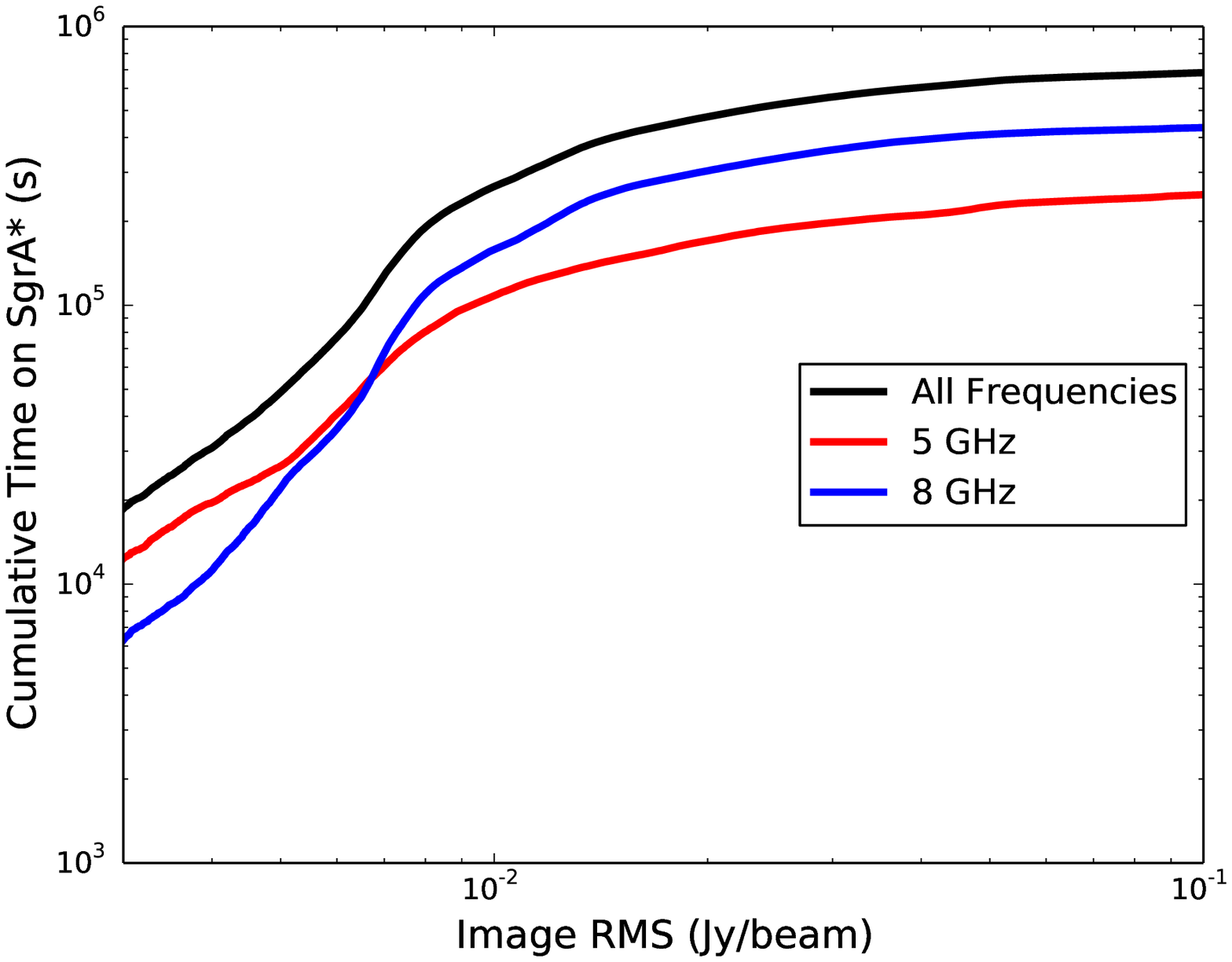}}
\caption{The cumulative integration time on \sgra\, from all images that have a lower rms than the abscissa value.}
\label{fig:timeonSGRA}
\end{figure}

\subsection{Transient Rate}
\label{ssec:transient_rate}

\begin{deluxetable}{cccc}
\tablewidth{0.47\textwidth}
\tablecolumns{4}
\tablecaption{Areas Surveyed and Time on \sgra\, for given Sensitivities}
\tablehead{   
  \colhead{Image RMS} &
  \colhead{Epochs$^{*}$} &
  \colhead{Area Surveyed} &
  \colhead{Time$^{\dagger}$ on \sgra} \\
  \colhead{(mJy)} &
  &
  \colhead{(deg$^2$)} &
  \colhead{(hr)}
}

\startdata
$\leq$\,5    &  4\,098 &  19.9 & 13.5  \\  
$\leq$\,10   & 25\,812 &  37.1 & 74.1  \\ 
$\leq$\,20   & 45\,713 &  73.6 & 131.7 \\ 
$\leq$\,50   & 59\,530 & 138.0 & 176.8 \\ 
$\leq$\,100  & 62\,602 & 166.5 & 189.8 \\ 
$\leq$\,200  & 64\,419 & 176.9 & 196.1 \\ 
$\leq$\,500  & 64\,856 & 178.6 & 197.4
\enddata
\tablecomments{\footnotesize Column 1 is the rms of an image; column 2 is the number of residual images that satisfy the image rms criteria; column 3 is the cumulative number of square degrees that an equivalent number of 10\,s images would produce; column 4 is the cumulative time on \sgra. \\
$^*$ An Epoch refers to a 10\,s, 20\,s, or 30\,s image, as determined by sample time of the visibility data.\\
$^\dagger$ The time on \sgra\, does not approach the stated 214 hours of observation time due to RFI flagging.}
\label{tab:rates}
\end{deluxetable}

The transient rate, $\rho$, which is just the number of detectable transients per 
unit solid angle and time, can be inferred from the observed number of transients 
in the analyzed VLA data set.  If we let $\Omega_i = \Omega_{A,i} \, \Delta t_i$ 
be the product of the solid angle ($\Omega_{A,i}$) and duration ($\Delta t_i$) of 
a single snapshot, then the total volume of the survey is just $\Omega = \sum_i \Omega_i$.
If we assume that the number of transients occuring in any given volume element is 
Poisson distributed, then the probability distribution for the number of observed
transients is given by

\begin{equation}
\label{eqn:p_counts}
P\left(n=k~|~\rho, \mathcal{I}\right) = \frac{ \left(\rho \Omega\right)^k e^{-\rho\Omega}}{k!}
\end{equation}
where $n$ is the number of transients detected in some volume $\Omega$, 
$\rho$ is the transient rate, and $\mathcal{I}$ encapsulates all other prior 
information (including the fact that the counts follow a Poisson distribution).  Using Bayes' 
theorem, Equation~\ref{eqn:p_counts} can be inverted to find the probability 
distribution for the transient rate given the observed counts:

\begin{equation}
\label{eqn:bayes}
P(\rho~|~n=k, \mathcal{I}) = \frac{P(n=k~|~\rho, \mathcal{I}) P(\rho~|~\mathcal{I})}{ P(n=k~|~\mathcal{I})}.
\end{equation}
Adopting a uniform prior on the rate, so that $\rho \in \left[0, \rho_{\rm u}\right]$ with some large upper limit given by $\rho_{\rm u}$, we find that

\begin{equation}
\label{eqn:p_rate}
P(\rho~|~n=k, \mathcal{I}) = \frac{ \Omega\left(\rho \Omega\right)^k e^{-\rho\Omega}}{k!} \,.
\end{equation}

Though not expressed explicitly in Equation~\ref{eqn:p_rate}, the number of 
events observed (and thus the rate inferred) is dependent upon the event 
detection threshold. The detection threshold for snapshot image $i$ is roughly 
$S_{{\rm min}, i} = 7 \sigma_{{\rm rms}, i}$, where $\sigma_{{\rm rms}, i}$ is the 
rms noise level in the image. Since an event is only detectable if its peak flux density 
exceeds the detection threshold, it is necessary to calculate the rate as a function of 
$S_{\rm min}$. The above-threshold event rate $r(S_{\rm min})$ can be calculated with 
Equation~\ref{eqn:p_rate} using the number of events $n(S_{\rm min})$ in a volume 
$\Omega(S_{\rm min})$ for all snapshot images with $S_{{\rm min},i} \leq S_{\rm min}$.
A summary of the areas surveyed and time on \sgra for given sensitivities is provided in 
Table~\ref{tab:rates}, and the cumulative integration time on \sgra as a function of image rms is plotted
in Figure~\ref{fig:timeonSGRA}.

Given the observed values for $n(S_{\rm min})$ and $\Omega(S_{\rm min})$, 
we can infer $\rho(S_{\rm min})$ to be the maximum likelihood value  of 
Equation~\ref{eqn:p_rate} with an uncertainty set by the most compact range 
containing 95\% of the posterior distribution.
For a detection threshold of $S_{\rm min} = 94~\rm mJy$, corresponding to the 
highest threshold at which both events are detectable, we observe two events in 
$\Omega(94~\rm mJy) = 0.14~{\rm hr\,deg}^2$, giving a rate of 
$\rho(n{=}2,\, S_{\rm min} {=} 94~\rm mJy) = 14^{+32}_{-12}~{\rm hr}^{-1}\,{\rm deg}^{-2}$.
Similarly, for a detection threshold of $S_{\rm min} = 149~\rm mJy$, 
where only \Eone\ is detectable, we observe one event in 
$\Omega(149~\rm mJy) = 0.22~{\rm hr\,deg}^2$, giving a rate of 
$\rho(n{=}1,\, S_{\rm min} {=} 149~\rm mJy) = 5^{+17}_{-4.8}~{\rm hr}^{-1}\,{\rm deg}^{-2}$.

\begin{figure}[t]
\centering
\centerline{\includegraphics[scale = 0.45]{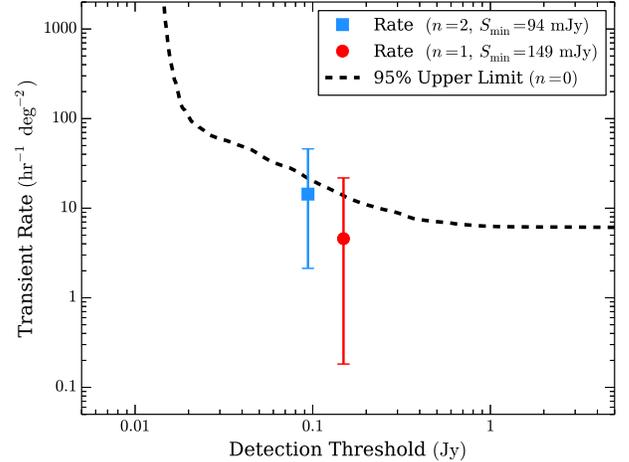}}
\caption{The transient rates as a function of detection threshold 
(7 times the image rms) with error bars corresponding to the most compact 95\% interval. 
The 95\% upper limits in the case of no detections is shown as a dashed line.  The point 
at 94~mJy corresponds to our detections of \Eone\ and \Etwo, as \Etwo\ occurred at 
94~mJy and \Eone\ (149~mJy) would have been detectable at that threshold. Similarly, 
the point at 149~mJy corresponds to the detection of \Eone.}
\label{fig:TimeRates}
\end{figure}

\begin{figure*}
\centering
\centerline{\includegraphics[width=0.75\textwidth]{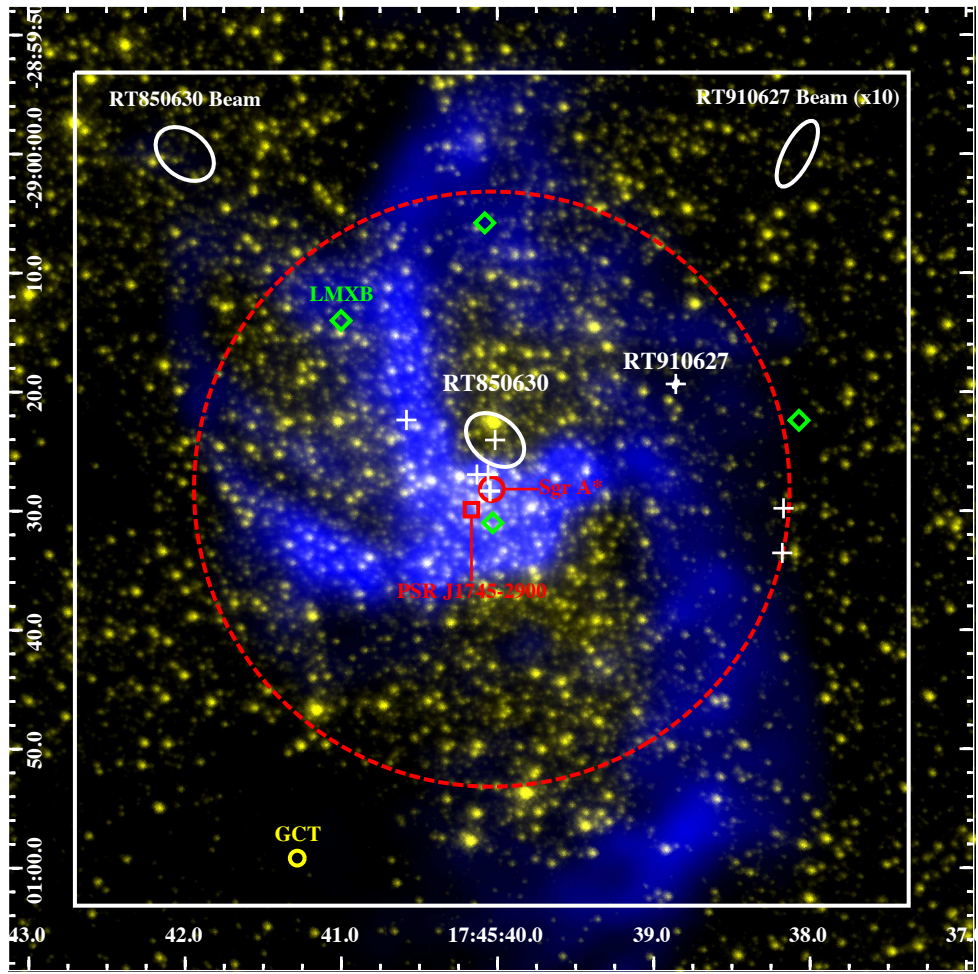}}
\caption{\footnotesize The candidate events overlaid as white crosses on a map of the Galactic center with 8~GHz VLA data in blue and Hubble~$1.9\mu\rm m$ data in yellow. The two most robust candidates are labeled with their transient identifier (\Eone\ and \Etwo). Candidate event 8 is located outside the field of view of this image. The white box is $70\arcsec \times 70\arcsec$ and the two red circles denote separations of $1\arcsec$ and $25\arcsec$ from \sgra.  The positions of the magnetar \gcmag, the GCT, and the LMXBs from \citet{mlb+05} are all indicated. The beam sizes and orientations for \Eone\ and \Etwo(x10) are shown at the top of the image.}
\label{fig:events}
\end{figure*}

Finally, since it is still possible that both of these events could be RFI 
and imaging artifacts, the 95\% upper limits for the rates with no detections 
are calculated and shown as a function of detection threshold in 
Figure~\ref{fig:TimeRates}.

\subsection{Galactic Center Rate}
\label{ssec:gc_rate}

If the transient events arise from a source population around \sgra\ 
that is entirely contained within the smallest image of our survey, then 
there is no longer any solid angle dependence on the rate. In such a case, 
the most natural rate is one of events per unit time, which we will call the 
Galactic center rate, $r$. Following Section~\ref{ssec:transient_rate}, the 
posterior for the Galactic center rate is 

\begin{equation}
\label{eqn:p_gcrate}
P(r~|~n=k, \mathcal{I}) = \frac{ T\left(r T\right)^k e^{-rT}}{k!} \,.
\end{equation}
where $n$ is the number of transients detected in some time $T$.

For a detection threshold of $S_{\rm min} = 94~\rm mJy$, we observe two events in 
$T(94~\rm mJy) = 4.30~{\rm days}$, giving a rate of 
$r(n{=}2,\, S_{\rm min} {=} 94~\rm mJy) = 0.47^{+0.53}_{-0.27}~{\rm day}^{-1}$.
Similarly, for detection thresholds of $S_{\rm min} = 149~\rm mJy$, where only \Eone\
is detectable, we observe one event in 
$T(149~\rm mJy) = 5.66~{\rm days}$, giving a rate of 
$r(n{=}1,\, S_{\rm min} {=} 149~\rm mJy) = 0.18^{+0.67}_{-0.17}~{\rm day}^{-1}$.

\section{Astrophysical Source of Observed Events}
\label{sec:origin}

Transient radio signals can arise from both coherent and incoherent emission 
processes from a wide range of astrophysical objects including compact objects, 
stars, and planets \citep{clm04}.  Given the typical duration, flux density, 
and rate of our observed transient events, we can determine if they could 
feasibly originate from known astrophysical sources. 

\subsection{Coherent or Incoherent?}

Radio emission can arise from both coherent and incoherent processes.   
Coherent emission (e.g., pulsar emission) typically results in steep 
spectra and a high degree of polarization, whereas incoherent radio emission 
(e.g., synchrotron radiation from jets) often produces spectra that are flat 
and wide-band with only moderate polarization. Theoretical estimates set the 
maximum brightness temperature of a synchrotron source to be 
$T_{\rm B} \sim 10^{11-12}~\rm K$ \citep{kpt69, readhead94}, which we adopt 
as the upper limit for incoherent emission.  Any source with 
$T_{\rm B} \gtrsim 10^{12}~\rm K$ must then arise from a coherent emission process.

The brightness temperature of a pulse is

\begin{multline}
T_{\rm B} = 1.2 \times 10^{13}~{\rm K}~
                                    \left(\frac{S_{\nu}}{100~{\rm mJy}}\right)
                                    \left(\frac{D}{8.5~{\rm kpc}}\right)^2 \\
                          \times    \left(\frac{\nu}{5~{\rm GHz}}\right)^{-2}
                                    \left(\frac{\Delta t_{\rm p}}{100~{\rm s}}\right)^{-2},
\label{eqn:bright_e1}
\end{multline}
where $S_{\nu}$ is the peak flux density at frequency $\nu$, $\Delta t_{\rm p}$ is 
the characteristic width of the pulse, and $D$ is the distance to the source \citep{clm04}.

For events like \Eone\  and \Etwo\ 
($S_{\nu} \sim 100~{\rm mJy}$, $\Delta t_{\rm p} \sim 100~{\rm s}$), 
the brightness temperature exceeds the threshold for incoherent emission 
if the source is at the Galactic center. However, the brightness temperatures 
for both sources can be reduced to $T_{\rm B} \lesssim 10^{12}~\rm K$ if they 
are foreground sources located $D \lesssim 2.5~\rm kpc$ from the Sun. 

\subsection{Galactic or Extragalactic?}
Our survey was designed to target transient sources located near \sgra, but
it is possible that the observed events originate from extragalactic sources or from 
nearby Galactic sources along the line of sight. If the sources are extragalactic, 
then they should occur isotropically on the sky.  Extrapolating the inferred rate from 
Section~\ref{ssec:transient_rate} to the whole sky gives an event rate of 
$\rho_{\rm xgal}(S_{\rm min}{=} 94~\rm mJy) = 
          1.4^{+3.1}_{-1.2}\times 10^7~{\rm sky}^{-1}\,{\rm day}^{-1}$, 
which is incredibly high.  For comparison, Fast Radio Burts (FRBs), a class of 
bright ($S_{\rm pk} \sim 1~\rm Jy$) short duration ($\Delta t \sim 1~\rm ms$) 
radio transients of presumed extragalactic origin occur at a rate of 
$\rho_{\rm FRB} \lesssim 7\times 10^4~{\rm sky}^{-1}\,{\rm day}^{-1}$ 
\citep{lbb15}.  Even at the relatively unexplored timescales of 
$\Delta t \sim 100\,\rm s$, the existence of an undetected class of extragalactic 
radio transients occurring $\gtrsim 100$ times more frequently than FRBs seems 
unlikely.  For the remainder of this analysis, we will only consider Galactic sources.

The source of the transient events, if real, are almost certainly Galactic.  However,
without the identification of a counterpart at some other wavelength, it is 
very difficult to determine if they are located near \sgra\ or are foreground 
sources seen along the line of sight.  As a result, we will not exclude possible 
astrophysical sources on the basis of distance alone.

\subsection{Pulsar Emission}

The most obvious sources of radio transient emission are pulsars, which can exhibit 
variability over a wide range of timescales.  Individual pulses typically have durations of
${\sim}1~\mu{\rm s}$ to ${\sim}100~{\rm ms}$ and vary in amplitude from pulse to pulse with each rotation of the pulsar.  The emission can be sporadic as well, with nulling pulsars and 
RRATs sometimes emitting only a few observable pulses over many hours or days \citep{mll+06}.  
On longer timescales, the pulsar emission can be turned on and off for days or weeks 
at a time by large-scale changes to the magnetosphere \citep{klo06}, and an otherwise 
consistent pulsar may have its observed flux density modulated by refractive and 
diffractive scintillation in the interstellar medium \citep{ssh00}.  
Despite the diversity of variability, most pulsars would simply be too faint to account
for our observed transient events.  For all the pulsars in the ATNF Pulsar 
Catalog\footnote{\texttt{http://www.atnf.csiro.au/research/pulsar/psrcat}} \citep{mht+05} 
with reported flux densities and distances, the maximum 5~GHz flux density 
would only be about $1-10~{\rm mJy}$ if placed at the Galactic center.  
Though the transient events could not be caused by typical pulsar emission, 
it is possible that they could be the result of one or many giant pulses, like those 
observed in the Crab pulsar (B0531+21).

The Crab pulsar is a young ($\tau \approx 1~\rm kyr$) pulsar that frequently emits 
giant pulses, which are pulses with amplitudes at least ten times larger than the average.  
These pulses are incredibly bright, with observed peak flux densities as high as 2.2~MJy 
in less than $0.9~{\rm ns}$ \citep{he07}.  
Much work has been done characterizing the Crab giant pulses, which seem to follow
a power-law distribution in amplitude \citep{lcu+95, btk+08, mnl+11, mml+12}.  
Most studies report the index for the differential power-law amplitude distribution,
$n(S) \propto S^{-\gamma}$, which gives the number of pulses within some amplitude
bin.  Measurements of the index range from $2 \lesssim \gamma \lesssim 4$ for 
observing frequencies from $\nu_{\rm obs} = 0.1-4~{\rm GHz}$ 
\citep[][and reference therein]{mml+12}.  Since the intrinsic widths of 
Crab giant pulses can be smaller than the instrumental resolution, the observed 
peak flux density is often less than the intrinsic value.  To avoid such complications,
it is convenient to work with distributions of pulse fluence 
(often called the pulse energy), which is the time integrated flux density 
of the pulse.  An additional benefit of using the pulse fluence instead of 
the amplitude is that the fluence is unaffected by scattering or other 
pulse shape distortions.

\citet{mnl+11} determined the cumulative fluence distribution of Crab giant pulses
at 1.6~GHz and found that the rate of pulses with $E  > E_{\rm p}$ goes roughly as 
$R(E>E_{\rm p}) = R_0 (E_{\rm p} / E_0)^{-1.9}$ for 
$E_{\rm p}>5\times 10^{-3}~\rm Jy\,s$.
From their Figure~9, we estimate $R_0 \approx 3\times10^{-3}~\rm s^{-1}$ for 
$E_0 \approx 0.1~\rm Jy\,s$.
The fluence of our \Eone\ is 
$E_1 \sim \rm 0.1~Jy \times 100~s = 10~Jy\,s$.
Scaling this event with frequency and distance for comparison with the Crab gives

\begin{equation}
\label{eqn:fluence_scale}
E_{\rm c, \nu_0} = E_{\rm gc, \nu} \left(\frac{\nu_0}{\nu}\right)^{ \alpha}
                                                       \left(\frac{d_0}{d_{\rm gc}}\right)^{-2} 
\end{equation}
where $d_0 = 2~\rm kpc$ is the distance to the Crab pulsar \citep{t+73}, 
$\nu_0 = 1.6~\rm GHz$, $\alpha = -1.6$ is a typical pulsar spectral index \citep{lyl95}, 
and $E_{\rm gc, \nu}$, $\nu$, and $d_{\rm gc} = 8.5~\rm kpc$ are the 
fluence, frequency, and distance to a transient originating at the Galactic center.  
For \Eone\ ($E_1 \sim 10~\rm Jy\,s$, $\nu = 5~\rm GHz$),
the scaled fluence is $E_{\rm c, \nu_0, 1} \sim 10^3~\rm Jy\,s$.
Assuming the cumulative frequency relation
from \citet{mnl+11} extends to these fluences, the expected rate of giant pulses 
from the Crab with $E > 10^3~\rm Jy\,s$ 
is $R\approx 7 \times 10^{-12}~{\rm s}^{-1}$, or about once every 4800~years.  
One event in 200~observing hours requires a population of ${\sim}1$ Crab-like pulsar at 
$330~\rm pc$, ${\sim}1000$ Crab-like pulsars at $2~\rm kpc$, or ${\sim}2\times 10^5$
Crab-like pulsars in the Galactic center. Such a large population is easily 
ruled out by previous multiwavelength constraints \citep{wcc12}.

\subsection{Radio Flares from Dwarf Stars}
Transient radio emission at GHz frequencies has also been seen in flares from 
dwarf stars.  About 10\% of the ${\sim}100$ observed dwarf stars with spectral 
types above M7 have some sort of radio activity \citep{berger06}.  Most of the 
flares from late-type dwarf stars have peak 5~GHz flux densities around 
$S_{\rm pk} \lesssim 1~\rm mJy$ and durations of $\tau \gtrsim 10~{\rm min}$
\citep{berger02, berger06, had+08}.  The largest flare of this type was seen 
in the M8~dwarf DENIS\,1048$-$3956 and consisted of 5~minute long pulses at 
4.8~GHz and 8.6~GHz with peak flux densities of $S_{4.8} \approx 6~\rm mJy$ and 
$S_{8.6}\approx 30~{\rm mJy}$, respectively \citep{bp05}.  Several of the 
flaring ultracool dwarfs show ${\sim}100\%$ circularly polarized emission and 
are periodic on ${\sim} \rm hour$ timescales consistent with the rotational 
period of the star \citep{hbl+07, had+08}.  

M-class dwarfs with spectral type below M7 also exhibit radio flaring and are much more
active at other wavelengths than their later-type counterparts.  Large flares with 
${\sim}100\%$ circular polarization have been observed in both AD~Leonis and DO~Cephei 
with durations of $\tau \sim 1~{\rm min}$ and peak 5~GHz flux densities of 
$S_{\rm pk, 5} \gtrsim 100~{\rm mJy}$ \citep{wjk89, skz+01}.
Follow-up studies of AD~Leo with larger bandwidth and higher time resolution have found
that these flares are comprised of short ($\tau \sim 30~\rm ms$) subpulses with 
fractional bandwidths of $\Delta \nu / \nu \approx 5\%$ \citep{ob06, ob08}.

The flares from active M~dwarf stars like AD~Leonis are a good match to our two 
best radio transient events in both peak flux density and observed duration 
($S_{\rm pk} \sim 100~\rm mJy$, $\tau \sim 1~\rm min$).  Though AD~Leonis is one of 
the most luminous flaring M~dwarf stars, it can only produce flares up to  
$S_{\rm max} \sim 1~\rm Jy$ at a distance of $d=4.9~\rm pc$.  A similar star 
would result in flares comparable with our transient events 
($S_{\rm pk} \sim 100~\rm mJy$) out to a distance of only $d\approx 16~\rm pc$.  
Adopting a local stellar mass density of 
$\rho_{\odot} = 0.085\pm 0.010~M_{\odot}~{\rm pc}^{-3}$ \citep{mcmillan11}, 
the enclosed mass in our FOV ($\theta = 70\arcsec$) is  
$M_\star \approx 10^{-5} M_\odot \, (d / 16~\rm pc)^3$, which is significantly 
less than the typical M~dwarf mass of $M \approx 0.075-0.5~\rm M_{\odot}$ \citep{lg11}.
Thus, it is unlikely that an M~dwarf would be found along our line of sight to \sgra.
Furthermore, our rejection of circularly polarized detections means our detections
do not sharing the polarization properties of M~dwarf star flaring activity.

\subsection{\gcrt}
The Galactic center radio transient \gcrt\ (hereafter, GCRT) is an intermittently emitting
source that has been detected by multiple observatories at 330~MHz 
\citep{hlk+05, hlr+06, hrp+07}.  Other radio transients had been detected near \sgra,
but they were isolated events that lasted many months or years \citep{zrg+92, bry+05}.  
In contrast, the GCRT emits ${\sim} 1~\rm Jy$ pulses at 330~MHz that last for about 
10~minutes.  The pulses repeat with a 77~minute period, though the emission duty cycle is
only ${\sim} 10\%$ \citep{hlk+05, hlr+06}.  The spectral index ($S \propto \nu^{\alpha}$) 
of the source is $\alpha = -4 \pm 3$ \citep{hlr+06, rhp+10}, but at least 
one burst was observed with a much steeper spectral index of 
$\alpha = -13.5 \pm 3.0$ \citep{hrp+07}.  Though originally reported to have a very 
small polarized fraction, a recent reanalysis by \citet{rhp+10} found strong time-varying
circular polarization in one GCRT burst.  Follow-up searches in X-rays and near-infrared have 
produced no viable counterpart \citep{hlk+05, khr+08}.

The GCRT bursts are of comparable duration (${\sim}$minutes) and 
fluence ($E_{\rm p} \sim 10~\rm Jy\,s$ for $\alpha \approx -1.6$)
to the observed transient events in our VLA sample.  Like our candidate
events, the GCRT has no known counterpart at other wavelengths.
While a time-varying circular polarization and periodic nature
of the emission are not seen in our events, all but one of the bursts from the
GCRT have previously been detected with only 
upper limits on the circular polarization \citep{rhp+10}.  Although it is certainly possible that 
there exists a GCRT-like source responsible for our reported transient events, 
the evidence to support this proposition is ambiguous.

\subsection{X-Ray Binaries}

X-ray binaries (XRBs) are yet another potential source of transient radio emission.
The binaries are comprised of a neutron star or black hole and a low-mass 
(low-mass X-ray binaries; LMXBs) or high-mass (high-mass X-ray binaries; HMXBs) 
companion. Outflows from XRB jets can produce synchrotron radio emission, which
can be transient during state transitions \citep{fk01}.  The radio outbursts can reach
peak flux densities of $S_{\rm pk} \sim 1~\rm Jy$ (with relatively flat spectral indices)
and are highly correlated with X-ray emission \citep{fk01, gfp03}.  An XRB origin for 
radio transients in the Galactic center is particularly appealing because there is evidence
for an overabundance of LMXBs in the inner parsec around \sgra\ \citep{mpb+05}.  Furthermore,
an LMXB has already been linked to a long-duration (${\sim}100\,\rm days$) radio transient 
within $2\farcs6$ from \sgra\ \citep{bry+05, mlb+05}. 

Though XRBs have the appropriate energetics and are well-represented in the Galactic center,
the durations of known outbursts are much longer than the ${\sim}100~\rm s$ seen in our events.  
Typical XRB outbursts last for hundreds of days and even the shortest are still about a day
\citep{bdv99, hrh+00}.  In the absence of any other supporting evidence (e.g., coincident
X-ray emission), there is not much support for an XRB origin for our observed transient events.
We note, however, that few surveys have been sensitive to the transient timescales reported here.

\section{Conclusions}
\label{sec:conclusions}

Using over 200~hours of archival VLA data from 1985$-$2005 at 5~and~8~GHz,
we have conducted a thorough search for radio transients in the Galactic center 
on timescales down to 30\,s.  Out of 23 possible transient candidates identified
by our automated processing and detection pipeline, eight passed a series of
tests to filter out RFI and imaging artifacts.  Of these eight, two are identified 
as promising transient event candidates by their smooth lightcurves. Though we 
have carefully tried to eliminate false-positives from RFI or imaging artifacts
through a series of confirmation tests, it is still possible that our two detections
are spurious. In particular, the detection of \Etwo\, appears to be dependent on
self-calibration and choice of reduction package. If the events are real, the inferred transient rate is 
$\rho(n{=}2,\, S_{\rm min} {=} 94~\rm mJy) = 14^{+32}_{-12}~{\rm hr}^{-1}\,{\rm deg}^{-2}$.
If the sources of the events are restricted to the inner few parsecs around \sgra, 
then similar bursts should occur a few times per week.  

Investigations of the possible astrophysical origins of the radio transients 
have proven to be inconclusive.  Typical pulsar emission is at least an order
of magnitude too weak at 5~and~8~GHz to account for the 
${\sim}100~\rm mJy$ transient events.  Crab-like giant pulses could 
potentially reach the fluences required, but would be far too rare and 
would require a level of scattering that may not be present in the Galactic center.
Dwarf flare stars have large circular polarization fractions, which are incompatible 
with our apparently unpolarized transient events.
X-ray binaries, which would otherwise be an excellent source class given their
large radio bursts and overdensity near \sgra, exhibit transient behavior on 
much longer timescales than the ${\sim}1~\rm min$ durations of our observed events.

Some bursts from \gcrt~have previously been detected with comparable duration, fluence, and
only an upper limit on the circular polarization. However, the steep spectral index
and variability in the properties of prior GCRT bursts makes an association with our 
reported events more uncertain.
It remains possible that the transient events could come from a new class of radio
sources or an unusual manifestation of a known class.  This ambiguous result
using only archival radio data provides yet another example of the necessity of 
concurrent multiwavelength observations.

Future work will modify the processing and detection pipeline for use on
data collected with the fully upgraded Karl G. Jansky VLA.  Though our images
are limited by dynamic range, the greatly expanded available
bandwidths should improve the quality of snapshot images through increased
u-v coverage.  Additionally, the much larger fractional bandwidths will allow 
for more stringent tests to eliminate false-positives caused by sidelobes, providing more
robust detections.  Finally, analysis of data from 
modern coordinated multiwavelength observing campaigns of \sgra\  
will greatly improve our understanding of any new transient sources in the 
Galactic center.

\section{Acknowledgements}

The National Radio Astronomy Observatory is a facility of the 
National Science Foundation operated under cooperative agreement
by Associated Universities, Inc. This research has made use of the
SIMBAD database, operated at CDS, Strasbourg, France, data 
obtained from the Chandra Data Archive and
the Chandra Source Catalog, and software provided by the Chandra X-ray
Center (CXC) in the application packages CIAO, ChIPS, and Sherpa. We
acknowledge support from the National Science Foundation for this work
through grant AST-1008213 at Cornell. DLK and SDC are additionally 
supported by NSF grant AST-1412421. Part of this research was
carried out at the Jet Propulsion Laboratory, California Institute of
Technology, under a contract with the National Aeronautics and Space
Administration. We would like to thank an anonymous referee whose
insights and comments have greatly improved the quality of this paper.

\bibliography{transients}

\begin{thebibliography}{67}
\expandafter\ifx\csname natexlab\endcsname\relax\def\natexlab#1{#1}\fi

\bibitem[{{Becker} {et~al.}(1994){Becker}, {White}, {Helfand}, \&
  {Zoonematkermani}}]{bwh+94}
{Becker}, R.~H., {White}, R.~L., {Helfand}, D.~J., \& {Zoonematkermani}, S.
  1994, \apjs, 91, 347

\bibitem[{{Bell} {et~al.}(2011){Bell}, {Fender}, {Swinbank}, {Miller-Jones},
  {et~al.}}]{bfs+11}
{Bell}, M.~E., {Fender}, R.~P., {Swinbank}, J., {Miller-Jones}, J.~C.~A.,
  {et~al.} 2011, \mnras, 415, 2

\bibitem[{{Belloni} {et~al.}(1999){Belloni}, {Dieters}, {van den Ancker},
  {et~al.}}]{bdv99}
{Belloni}, T., {Dieters}, S., {van den Ancker}, M.~E., {et~al.} 1999, \apj,
  527, 345

\bibitem[{{Berger}(2002)}]{berger02}
{Berger}, E. 2002, \apj, 572, 503

\bibitem[{{Berger}(2006)}]{berger06}
---. 2006, \apj, 648, 629

\bibitem[{{Berger} {et~al.}(2001){Berger}, {Ball}, {Becker}, {Clarke},
  {et~al.}}]{bbb+01}
{Berger}, E., {Ball}, S., {Becker}, K.~M., {Clarke}, M., {et~al.} 2001, \nat,
  410, 338

\bibitem[{{Bhat} {et~al.}(2008){Bhat}, {Tingay}, \& {Knight}}]{btk+08}
{Bhat}, N.~D.~R., {Tingay}, S.~J., \& {Knight}, H.~S. 2008, \apj, 676, 1200

\bibitem[{{Bower} {et~al.}(2005){Bower}, {Roberts}, {Yusef-Zadeh}, {Backer},
  {et~al.}}]{bry+05}
{Bower}, G.~C., {Roberts}, D.~A., {Yusef-Zadeh}, F., {Backer}, D.~C., {et~al.}
  2005, \apj, 633, 218

\bibitem[{{Bower} \& {Saul}(2011)}]{bs+11}
{Bower}, G.~C., \& {Saul}, D. 2011, \apjl, 728, L14

\bibitem[{{Bower} {et~al.}(2007){Bower}, {Saul}, {Bloom}, {Bolatto},
  {Filippenko}, {Foley}, \& {Perley}}]{bsb+07}
{Bower}, G.~C., {Saul}, D., {Bloom}, J.~S., {Bolatto}, A., {Filippenko}, A.~V.,
  {Foley}, R.~J., \& {Perley}, D. 2007, \apj, 666, 346

\bibitem[{{Bridle} \& {Schwab}(1999)}]{bs+99}
{Bridle}, A.~H., \& {Schwab}, F.~R. 1999, in Astronomical Society of the
  Pacific Conference Series, Vol. 180, Synthesis Imaging in Radio Astronomy II,
  ed. G.~B. {Taylor}, C.~L. {Carilli}, \& R.~A. {Perley}, 371

\bibitem[{{Burgasser} \& {Putman}(2005)}]{bp05}
{Burgasser}, A.~J., \& {Putman}, M.~E. 2005, \apj, 626, 486

\bibitem[{{Chatterjee} {et~al.}(2005){Chatterjee}, {Goss}, \&
  {Brisken}}]{cgb+05}
{Chatterjee}, S., {Goss}, W.~M., \& {Brisken}, W.~F. 2005, \apjl, 634, L101

\bibitem[{{Cordes} {et~al.}(2004){Cordes}, {Lazio}, \& {McLaughlin}}]{clm04}
{Cordes}, J.~M., {Lazio}, T.~J.~W., \& {McLaughlin}, M.~A. 2004, nar, 48, 1459

\bibitem[{{Eatough} {et~al.}(2013){Eatough}, {Falcke}, {Karuppusamy},
  {et~al.}}]{efk+13}
{Eatough}, R.~P., {Falcke}, H., {Karuppusamy}, R., {et~al.} 2013, \nat, 501,
  391

\bibitem[{{Evans} {et~al.}(1981){Evans}, {Fenimore}, {Klebesadel},
  {et~al.}}]{efk+81}
{Evans}, W.~D., {Fenimore}, E.~E., {Klebesadel}, R.~W., {et~al.} 1981, \apss,
  75, 35

\bibitem[{{Fender} \& {Kuulkers}(2001)}]{fk01}
{Fender}, R.~P., \& {Kuulkers}, E. 2001, \mnras, 324, 923

\bibitem[{{Frail} {et~al.}(2012){Frail}, {Kulkarni}, {Ofek}, {Bower}, \&
  {Nakar}}]{fko12}
{Frail}, D.~A., {Kulkarni}, S.~R., {Ofek}, E.~O., {Bower}, G.~C., \& {Nakar},
  E. 2012, \apj, 747, 70

\bibitem[{{Gallo} {et~al.}(2003){Gallo}, {Fender}, \& {Pooley}}]{gfp03}
{Gallo}, E., {Fender}, R.~P., \& {Pooley}, G.~G. 2003, \mnras, 344, 60

\bibitem[{{Gillessen} {et~al.}(2013){Gillessen}, {Genzel}, {Fritz},
  {et~al.}}]{ggf+13}
{Gillessen}, S., {Genzel}, R., {Fritz}, T.~K., {et~al.} 2013, \apj, 763, 78

\bibitem[{{Hallinan} {et~al.}(2008){Hallinan}, {Antonova}, {Doyle},
  {et~al.}}]{had+08}
{Hallinan}, G., {Antonova}, A., {Doyle}, J.~G., {et~al.} 2008, \apj, 684, 644

\bibitem[{{Hallinan} {et~al.}(2007){Hallinan}, {Bourke}, {Lane}, {Antonova},
  {Zavala}, {Brisken}, {Boyle}, {Vrba}, {Doyle}, \& {Golden}}]{hbl+07}
{Hallinan}, G., {Bourke}, S., {Lane}, C., {Antonova}, A., {Zavala}, R.~T.,
  {Brisken}, W.~F., {Boyle}, R.~P., {Vrba}, F.~J., {Doyle}, J.~G., \& {Golden},
  A. 2007, \apjl, 663, L25

\bibitem[{{Hankins} \& {Eilek}(2007)}]{he07}
{Hankins}, T.~H., \& {Eilek}, J.~A. 2007, \apj, 670, 693

\bibitem[{{Hjellming} {et~al.}(2000){Hjellming}, {Rupen}, {Hunstead},
  {et~al.}}]{hrh+00}
{Hjellming}, R.~M., {Rupen}, M.~P., {Hunstead}, R.~W., {et~al.} 2000, \apj,
  544, 977

\bibitem[{{Hurley} {et~al.}(2000){Hurley}, {Lund}, {Brandt}, {Barat},
  {et~al.}}]{hbb+00}
{Hurley}, K., {Lund}, N., {Brandt}, S., {Barat}, C., {et~al.} 2000, \apjs, 128,
  549

\bibitem[{{Hyman} {et~al.}(2005){Hyman}, {Lazio}, {Kassim}, {Ray}, {Markwardt},
  \& {Yusef-Zadeh}}]{hlk+05}
{Hyman}, S.~D., {Lazio}, T.~J.~W., {Kassim}, N.~E., {Ray}, P.~S., {Markwardt},
  C.~B., \& {Yusef-Zadeh}, F. 2005, \nat, 434, 50

\bibitem[{{Hyman} {et~al.}(2006){Hyman}, {Lazio}, {Roy}, {et~al.}}]{hlr+06}
{Hyman}, S.~D., {Lazio}, T.~J.~W., {Roy}, S., {et~al.} 2006, \apj, 639, 348

\bibitem[{{Hyman} {et~al.}(2007){Hyman}, {Roy}, {Pal}, {et~al.}}]{hrp+07}
{Hyman}, S.~D., {Roy}, S., {Pal}, S., {et~al.} 2007, \apjl, 660, L121

\bibitem[{{Hyman} {et~al.}(2009){Hyman}, {Wijnands}, {Lazio}, {Pal},
  {et~al.}}]{hwl+09}
{Hyman}, S.~D., {Wijnands}, R., {Lazio}, T.~J.~W., {Pal}, S., {et~al.} 2009,
  \apj, 696, 280

\bibitem[{{Kaplan} {et~al.}(2008){Kaplan}, {Hyman}, {Roy}, {Bandyopadhyay},
  {Chakrabarty}, {Kassim}, {Lazio}, \& {Ray}}]{khr+08}
{Kaplan}, D.~L., {Hyman}, S.~D., {Roy}, S., {Bandyopadhyay}, R.~M.,
  {Chakrabarty}, D., {Kassim}, N.~E., {Lazio}, T.~J.~W., \& {Ray}, P.~S. 2008,
  \apj, 687, 262

\bibitem[{{Kedziora-Chudczer} {et~al.}(2001){Kedziora-Chudczer}, {Jauncey},
  {Wieringa}, {et~al.}}]{kjw01}
{Kedziora-Chudczer}, L.~L., {Jauncey}, D.~L., {Wieringa}, M.~H., {et~al.} 2001,
  \mnras, 325, 1411

\bibitem[{{Kellermann} \& {Pauliny-Toth}(1969)}]{kpt69}
{Kellermann}, K.~I., \& {Pauliny-Toth}, I.~I.~K. 1969, \apjl, 155, L71

\bibitem[{{Klebesadel} {et~al.}(1980){Klebesadel}, {Evans}, {Glore},
  {et~al.}}]{keg+80}
{Klebesadel}, R.~W., {Evans}, W.~D., {Glore}, J.~P., {et~al.} 1980, IEEE
  Transactions on Geoscience and Remote Sensing, 18, 76

\bibitem[{{Kramer} {et~al.}(2006){Kramer}, {Lyne}, {O'Brien}, {et~al.}}]{klo06}
{Kramer}, M., {Lyne}, A.~G., {O'Brien}, J.~T., {et~al.} 2006, Science, 312, 549

\bibitem[{{Law} {et~al.}(2015){Law}, {Bower}, {Burke-Spolaor},
  {et~al.}}]{lbb15}
{Law}, C.~J., {Bower}, G.~C., {Burke-Spolaor}, S., {et~al.} 2015, \apj, 807, 16

\bibitem[{{Lazio} \& {Cordes}(1998{\natexlab{a}})}]{lc+98}
{Lazio}, T.~J.~W., \& {Cordes}, J.~M. 1998{\natexlab{a}}, \apjs, 118, 201

\bibitem[{{Lazio} \& {Cordes}(1998{\natexlab{b}})}]{lc98}
---. 1998{\natexlab{b}}, \apj, 505, 715

\bibitem[{{L{\'e}pine} \& {Gaidos}(2011)}]{lg11}
{L{\'e}pine}, S., \& {Gaidos}, E. 2011, \aj, 142, 138

\bibitem[{{Lorimer} {et~al.}(1995){Lorimer}, {Yates}, {Lyne}, \&
  {Gould}}]{lyl95}
{Lorimer}, D.~R., {Yates}, J.~A., {Lyne}, A.~G., \& {Gould}, D.~M. 1995,
  \mnras, 273, 411

\bibitem[{{Lundgren} {et~al.}(1995){Lundgren}, {Cordes}, {Ulmer}, {Matz},
  {et~al.}}]{lcu+95}
{Lundgren}, S.~C., {Cordes}, J.~M., {Ulmer}, M., {Matz}, S.~M., {et~al.} 1995,
  \apj, 453, 433

\bibitem[{{Majid} {et~al.}(2011){Majid}, {Naudet}, {Lowe}, \&
  {Kuiper}}]{mnl+11}
{Majid}, W.~A., {Naudet}, C.~J., {Lowe}, S.~T., \& {Kuiper}, T.~B.~H. 2011,
  \apj, 741, 53

\bibitem[{{Manchester} {et~al.}(2005){Manchester}, {Hobbs}, {Teoh}, \&
  {Hobbs}}]{mht+05}
{Manchester}, R.~N., {Hobbs}, G.~B., {Teoh}, A., \& {Hobbs}, M. 2005, \aj, 129,
  1993

\bibitem[{{McLaughlin} {et~al.}(2006){McLaughlin}, {Lyne}, {Lorimer}, {Kramer},
  {et~al.}}]{mll+06}
{McLaughlin}, M.~A., {Lyne}, A.~G., {Lorimer}, D.~R., {Kramer}, M., {et~al.}
  2006, \nat, 439, 817

\bibitem[{{McMillan}(2011)}]{mcmillan11}
{McMillan}, P.~J. 2011, \mnras, 414, 2446

\bibitem[{{McMullin} {et~al.}(2007){McMullin}, {Waters}, {Schiebel}, {Young},
  \& {Golap}}]{mws+07}
{McMullin}, J.~P., {Waters}, B., {Schiebel}, D., {Young}, W., \& {Golap}, K.
  2007, in Astronomical Society of the Pacific Conference Series, Vol. 376,
  Astronomical Data Analysis Software and Systems XVI, ed. R.~A. {Shaw},
  F.~{Hill}, \& D.~J. {Bell}, 127

\bibitem[{{Mickaliger} {et~al.}(2012){Mickaliger}, {McLaughlin}, {Lorimer},
  {Langston}, {et~al.}}]{mml+12}
{Mickaliger}, M.~B., {McLaughlin}, M.~A., {Lorimer}, D.~R., {Langston}, G.~I.,
  {et~al.} 2012, \apj, 760, 64

\bibitem[{{Miralda-Escud{\'e}} \& {Gould}(2000)}]{meg00}
{Miralda-Escud{\'e}}, J., \& {Gould}, A. 2000, \apj, 545, 847

\bibitem[{{Mooley} {et~al.}(2013){Mooley}, {Frail}, {Ofek}, {Miller},
  {Kulkarni}, \& {Horesh}}]{mfo+13}
{Mooley}, K.~P., {Frail}, D.~A., {Ofek}, E.~O., {Miller}, N.~A., {Kulkarni},
  S.~R., \& {Horesh}, A. 2013, \apj, 768, 165

\bibitem[{{Mori} {et~al.}(2013){Mori}, {Gotthelf}, {Zhang}, {An},
  {et~al.}}]{mgz+13}
{Mori}, K., {Gotthelf}, E.~V., {Zhang}, S., {An}, H., {et~al.} 2013, \apjl,
  770, L23

\bibitem[{{Muno} {et~al.}(2009){Muno}, {Bauer}, {Baganoff}, {et~al.}}]{mbb+09}
{Muno}, M.~P., {Bauer}, F.~E., {Baganoff}, F.~K., {et~al.} 2009, \apjs, 181,
  110

\bibitem[{{Muno} {et~al.}(2005{\natexlab{a}}){Muno}, {Lu}, {Baganoff},
  {et~al.}}]{mlb+05}
{Muno}, M.~P., {Lu}, J.~R., {Baganoff}, F.~K., {et~al.} 2005{\natexlab{a}},
  \apj, 633, 228

\bibitem[{{Muno} {et~al.}(2005{\natexlab{b}}){Muno}, {Pfahl}, {Baganoff},
  {et~al.}}]{mpb+05}
{Muno}, M.~P., {Pfahl}, E., {Baganoff}, F.~K., {et~al.} 2005{\natexlab{b}},
  \apjl, 622, L113

\bibitem[{{Osten} \& {Bastian}(2006)}]{ob06}
{Osten}, R.~A., \& {Bastian}, T.~S. 2006, \apj, 637, 1016

\bibitem[{{Osten} \& {Bastian}(2008)}]{ob08}
---. 2008, \apj, 674, 1078

\bibitem[{{Readhead}(1994)}]{readhead94}
{Readhead}, A.~C.~S. 1994, \apj, 426, 51

\bibitem[{{Roy} {et~al.}(2010){Roy}, {Hyman}, {Pal}, {et~al.}}]{rhp+10}
{Roy}, S., {Hyman}, S.~D., {Pal}, S., {et~al.} 2010, \apjl, 712, L5

\bibitem[{{Sch{\"o}del} {et~al.}(2009){Sch{\"o}del}, {Merritt}, \&
  {Eckart}}]{sme+09}
{Sch{\"o}del}, R., {Merritt}, D., \& {Eckart}, A. 2009, \aap, 502, 91

\bibitem[{{Schwab}(1984)}]{schwab84}
{Schwab}, F.~R. 1984, \aj, 89, 1076

\bibitem[{{Shannon} \& {Johnston}(2013)}]{sj13}
{Shannon}, R.~M., \& {Johnston}, S. 2013, \mnras, 435, L29

\bibitem[{{Stepanov} {et~al.}(2001){Stepanov}, {Kliem}, {Zaitsev},
  {et~al.}}]{skz+01}
{Stepanov}, A.~V., {Kliem}, B., {Zaitsev}, V.~V., {et~al.} 2001, \aap, 374,
  1072

\bibitem[{{Stinebring} {et~al.}(2000){Stinebring}, {Smirnova}, {Hankins},
  {et~al.}}]{ssh00}
{Stinebring}, D.~R., {Smirnova}, T.~V., {Hankins}, T.~H., {et~al.} 2000, \apj,
  539, 300

\bibitem[{{Thyagarajan} {et~al.}(2011){Thyagarajan}, {Helfand}, {White}, \&
  {Becker}}]{thw+11}
{Thyagarajan}, N., {Helfand}, D.~J., {White}, R.~L., \& {Becker}, R.~H. 2011,
  \apj, 742, 49

\bibitem[{{Trimble}(1973)}]{t+73}
{Trimble}, V. 1973, \pasp, 85, 579

\bibitem[{{Wharton} {et~al.}(2012){Wharton}, {Chatterjee}, {Cordes},
  {et~al.}}]{wcc12}
{Wharton}, R.~S., {Chatterjee}, S., {Cordes}, J.~M., {et~al.} 2012, \apj, 753,
  108

\bibitem[{{White} {et~al.}(1989){White}, {Jackson}, \& {Kundu}}]{wjk89}
{White}, S.~M., {Jackson}, P.~D., \& {Kundu}, M.~R. 1989, \apjs, 71, 895

\bibitem[{{Williams} {et~al.}(2013){Williams}, {Berger}, \&
  {Zauderer}}]{wbz+13}
{Williams}, P.~K.~G., {Berger}, E., \& {Zauderer}, B.~A. 2013, \apjl, 767, L30

\bibitem[{{Zhao} {et~al.}(1992){Zhao}, {Roberts}, {Goss}, {Frail},
  {et~al.}}]{zrg+92}
{Zhao}, J.-H., {Roberts}, D.~A., {Goss}, W.~M., {Frail}, D.~A., {et~al.} 1992,
  Science, 255, 1538

\end{thebibliography}

\end{document}